# Warm, Water-Depleted Rocky Exoplanets with Surface Ionic Liquids: A Proposed Class for Planetary Habitability

Rachana Agrawal[1], Sara Seager[1,2,3,4,*], Iaroslav Iakubivskyi[1,5], Weston P. Buchanan[1], Ana Glidden[1,4], Maxwell D. Seager[6], William Bains[7], Jingcheng Huang[1], Janusz J. Petkowski[8,9,*]

[1] Department of Earth, Atmospheric and Planetary Sciences, [2] Department of Physics, Massachusetts Institute of
[3] Department of Aeronautical and Astronautical Engineering, [4] Kavli Institute for Astrophysics and Space Research, Massachusetts Institute of Technology, 77 Massachusetts Avenue, Cambridge, MA 02139, USA
[5] Tartu Observatory, University of Tartu, 1 Observatooriumi, 61602 Tõravere, Estonia
[6] Department of Chemistry and Biochemistry, Worcester Polytechnic Institute, Worcester, MA 01609, USA
[7] School of Physics and Astronomy, Cardiff University, 4 The Parade, Cardiff CF24 3AA, UK
[8] Faculty of Environmental Engineering, Wroclaw University of Science and Technology, 50-370 Wroclaw, Poland
[9] JJ Scientific, Warsaw, Mazowieckie, Poland

*Corresponding author: Sara Seager, Janusz J. Petkowski
Email:  seager@mit.edu, janusz.petkowski@pwr.edu.pl

**Abstract** The discovery of thousands of exoplanets and the emergence of telescopes capable of exoplanet atmospheric characterization have intensified the search for habitable worlds. Due to selection biases, many exoplanets under study are planets deemed inhospitable because their surfaces are too warm to support liquid water. We propose that such planets could still support life through ionic liquids: liquid salts with negligible vapor pressure that can persist on warm planets with thin atmospheres, where liquid water cannot. Ionic liquids have not previously been considered as naturally occurring substances, and thus have not been discussed in planetary science. We demonstrate in laboratory experiments that ionic liquids can form from planetary materials: sulfuric acid combined with nitrogen-containing organic molecules. Sulfuric acid can be volcanic in origin, and organic compounds are commonly found on planetary bodies. The required planetary surface is water-depleted and must support sulfuric acid transiently in liquid phase to dissolve organics, followed by evaporation of excess liquid—conditions spanning approximately 300 K at $10^{-7}$ atm to 350-470 K at 0.01 atm. Because ionic liquids have extremely low vapor pressures, they are not prone to evaporation, allowing small droplets or pools to persist without ocean-like reservoirs. Ionic liquids' miniscule vapor pressure at room temperature suggests possible stability on planets with negligible atmospheres, shielded by magnetic fields or rock crevices against harsh cosmic radiation. Ionic liquids can stably dissolve enzymes and other biomolecules, enabling biocatalysis and offering a plausible solvent for life—broadening the definition of habitable worlds.

**Significance Statement** The search for habitable exoplanets has intensified with new telescopes and a growing number of exoplanets. Yet, many known exoplanets are too warm for surface liquid water and therefore considered inhospitable to life. Liquid is a fundamental requirement for life as we understand it, but whether that liquid has to be water is not known. We propose that such planets could still support life through ionic liquids, substances with negligible vapor pressure that remain liquid in warm, low-pressure conditions, even approaching vacuum. Our experiments show that ionic liquids can form from planetary materials—sulfuric acid and nitrogen-containing organic compounds—offering a potential pathway for life on warm, thin-atmosphere, water-depleted worlds.

## 1. Introduction

Humanity's understanding of habitable worlds has expanded dramatically over the past few decades. We now know that several of Jupiter and Saturn's icy moons, including Europa, show strong evidence of subsurface, salty global oceans (1), with the Europa Clipper (2) and Jupiter Icy Moons Explorer missions (3) now en route to explore Europa further. Titan, the only solar system body other than Earth with persistent surface liquids, has methane and ethane lakes. The Dragonfly mission is currently under development to explore Titan's surface and atmosphere (4). Additionally, Venus has liquid droplets of sulfuric acid in its cloud layers, and new laboratory research shows that a subset of biologically-relevant molecules remain stable in concentrated sulfuric acid, opening new doors to considering life in the temperate layers of the Venus atmosphere (5–12). With thousands of exoplanets now known, we have growing array of possible habitable worlds, given their incredible diversity in mass, radius, and orbit. Our ability to fully observe and characterize terrestrial exoplanets with Earth-like conditions is still limited by current technology, motivating us to explore non-traditional pathways to habitability.

The model of terrestrial life, based on liquid water is so well established that considerations of habitability have been reserved largely for those rocky exoplanets that could have large stable surface reservoirs of liquid water and a supporting atmosphere. However, the diversity of environments in our solar system and the growing catalog of exoplanets inspire us to consider life in settings very different from Earth's. Therefore, we adopt the more general set of life's base requirements: a liquid solvent, temperatures suitable for covalent bonds (so complex molecules can form), and an energy source (13).

Under these general requirements we propose a new category of habitable planets—inspired by the observation that ionic liquids remain liquid even in vacuum at room temperature, and can form from common planetary chemicals. Ionic liquids enable warm, rocky super Earths without surface water and thin atmospheres to have surface liquid, and hence potential to be habitable.

Ionic liquids are salts in a liquid state, typically composed of ions rather than neutral molecules. They can remain liquid at a wide range of temperatures and pressures, including conditions where water would evaporate or freeze. Ionic liquids are formally defined as salts with melting temperatures below 100 °C (14). Ionic liquids differ from conventional liquids in that they typically lack a triple point. Rather than boiling, they decompose at the gas-phase boundary, and they often crystallize at temperatures above their melting points. Ionic liquid phase behavior is thus governed by solid-liquid equilibria and, in some cases, glass transitions, rather than solid-liquid-gas coexistence. Ionic liquids differ from each other in their characteristics (decomposition temperature, viscosity, etc.) but all share the property of very low vapor pressures. For example, many ionic liquids have vapor pressures as low as $10^{-15}$ bar at 300 K, remaining effective non-volatile under vacuum conditions (15).

There are thousands of known ionic liquids, with millions likely awaiting discovery (e.g., (16, 17)). Ionic liquids have a wide range of applications, including electrochemical (e.g., in batteries), industrial catalysis, biopolymer processing, and pharmacology (18). Ionic liquids are synthetic[†]

---

[†] The only known example of the natural formation of ionic liquids on Earth is a result of detoxification of a venom of the fire ant (*Solenopsis invicta*) by a rival ant species, the tawny crazy ant (*Nylanderia fulva*) (46). The fire ant

and have not previously been considered as naturally occurring substances, and so have not been discussed in the context of planetary science.

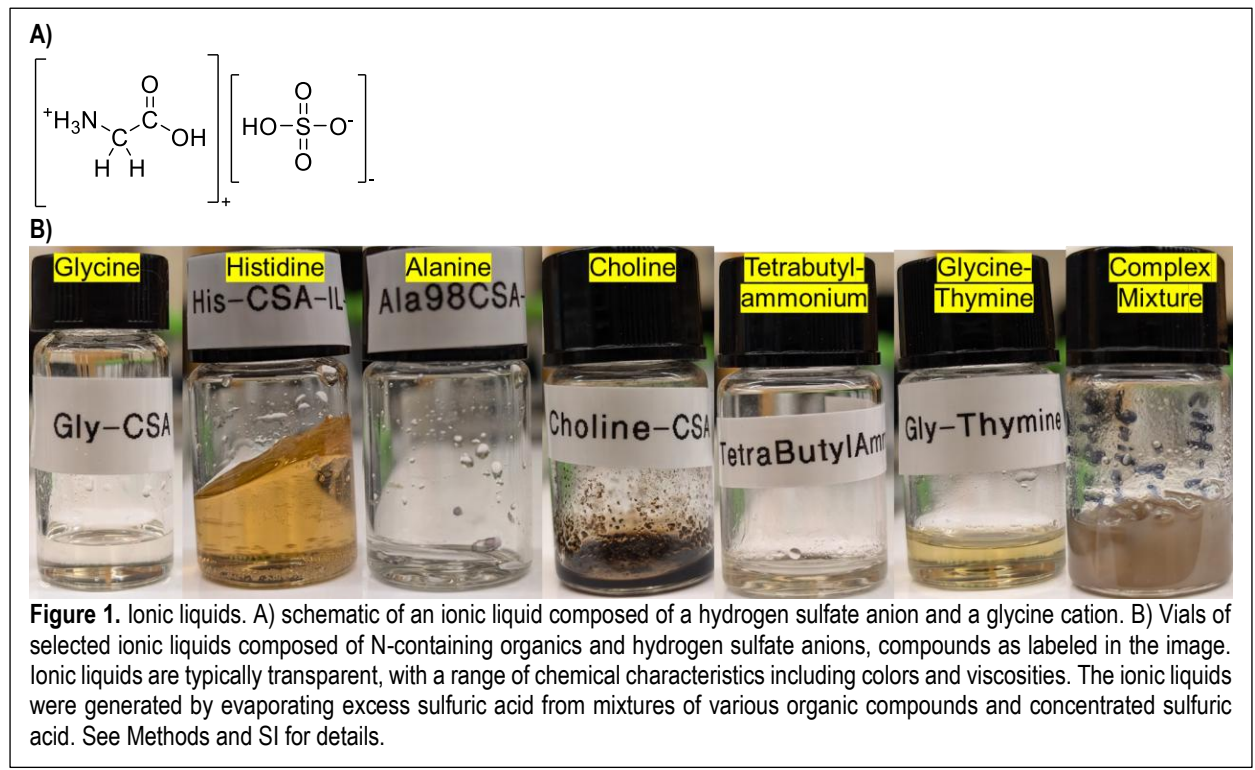

**Figure 1.** Ionic liquids. A) schematic of an ionic liquid composed of a hydrogen sulfate anion and a glycine cation. B) Vials of selected ionic liquids composed of N-containing organics and hydrogen sulfate anions, compounds as labeled in the image. Ionic liquids are typically transparent, with a range of chemical characteristics including colors and viscosities. The ionic liquids were generated by evaporating excess sulfuric acid from mixtures of various organic compounds and concentrated sulfuric acid. See Methods and SI for details.

## 2. Results

We have found that ionic liquids form from planetary materials: concentrated sulfuric acid combined with a wide variety of nitrogen-containing organic molecules (Figure 1) under a range of temperature-pressure conditions (see Methods). The ionic liquid is composed of a hydrogen sulfate ($HSO_4^-$) anionic component and a protonated nitrogen-containing organic compound cationic component (19) (hydrogen sulfate ionic liquids; Figure 1). In terms of planetary materials, the $HSO_4^-$ anion comes from sulfuric acid and the cationic component originates from organics found in meteorites, asteroid surfaces, and elsewhere, and also considered in prebiotic chemistry scenarios (20).

We form ionic liquids by first dissolving organic compounds in concentrated sulfuric acid followed by evaporation of the excess concentrated sulfuric acid under low pressure (at $10^{-5}$ bar and warm temperatures (100+/-5 °C)). For the evaporation procedure we use a custom-built in-house vacuum

---

produces a deadly solenopsin insecticide containing a piperidine ring. After being sprayed with fire ant solenopsin venom, the tawny crazy ant can detoxify by grooming itself with its own venom, formic acid. The formic acid mixes with the *S. invicta* venom and protonates the N atoms of the piperidine rings. The resulting mixture is a viscous naturally produced ionic liquid that is no longer toxic to *N. fulva*.

chamber. Other combinations of temperatures and pressures also form ionic liquids (see Methods and SI, Section S2).

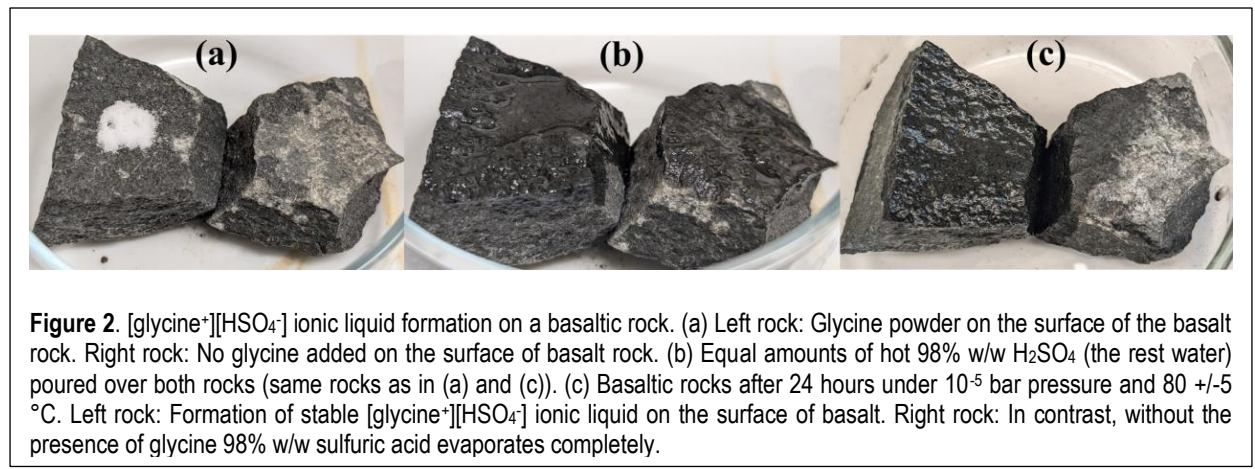

**Figure 2**. [glycine+][HSO4-] ionic liquid formation on a basaltic rock. (a) Left rock: Glycine powder on the surface of the basalt rock. Right rock: No glycine added on the surface of basalt rock. (b) Equal amounts of hot 98% w/w $H_2SO_4$ (the rest water) poured over both rocks (same rocks as in (a) and (c)). (c) Basaltic rocks after 24 hours under $10^{-5}$ bar pressure and 80 +/-5 °C. Left rock: Formation of stable [glycine+][HSO4-] ionic liquid on the surface of basalt. Right rock: In contrast, without the presence of glycine 98% w/w sulfuric acid evaporates completely.

We find ionic liquid formation for a wide variety of organic molecules that contain a nitrogen atom, including amino acids, aliphatic amines, nucleic acid bases, and other aromatic heterocycles. Once dissolved in concentrated sulfuric acid the N-containing organic molecules become protonated and gain a stable positive charge. This means that any nitrogen-containing organic molecule that is stable and soluble in concentrated sulfuric acid could form ionic liquids upon evaporation of the excess sulfuric acid.

We highlight ionic liquids with glycine, histidine, alanine, choline, tetrabutylammonium, as well as a mixture of glycine and thymine, and also a mixture of glycine, thymine, stearic acid, paraffin, and naphthalene (Figure 1).

We find that as long as the positively charged organics are present, ionic liquids can also form from complex mixtures of diverse compounds, including hydrocarbons, carboxylic acids, sugars (saccharose), and other chemicals (see SI, Table S1). The mixture of compounds indicates that impure, reactive and highly complex solutions of organics do not prevent the formation of ionic liquids (see SI Section S3 Figure S8). Furthermore: the concentration of sulfuric acid (i.e., dilution in water); the concentration of organics in sulfuric acid; and the overall volume of the initial organics and concentrated sulfuric acid mixture (down to the nL volume we could measure) do not inhibit the formation of ionic liquid, (see SI, Section S4).

While we do not expect ionic liquids to be pure substances (i.e., formed from a single type of organic molecule) here we largely explore individual ionic liquids for proper chemical characterization, while also including two example mixtures.

For more realistic planetary-like conditions we used basaltic rock as a platform to form ionic liquids. We used a mixture of N-containing organics and concentrated sulfuric acid on the surface of basaltic rocks, under a variety of conditions (80 °C and $10^{-5}$ bar, as well as room temperature, room pressure both in dry air and room humidity). The ionic liquids that form on the surface of basaltic rocks are stable to further reactivity (Figure 2; see SI, Section S4, Figures S18-S19).

We support the composition of the $HSO_4^-$ ionic liquids with Fourier transform infrared spectroscopy (FTIR) and $^1H$ nuclear magnetic resonance (NMR) spectroscopy. We use [glycine$^+$][HSO$_4^-$] as our main example with several others detailed in the SI (Table S1 and SI, Section S3).

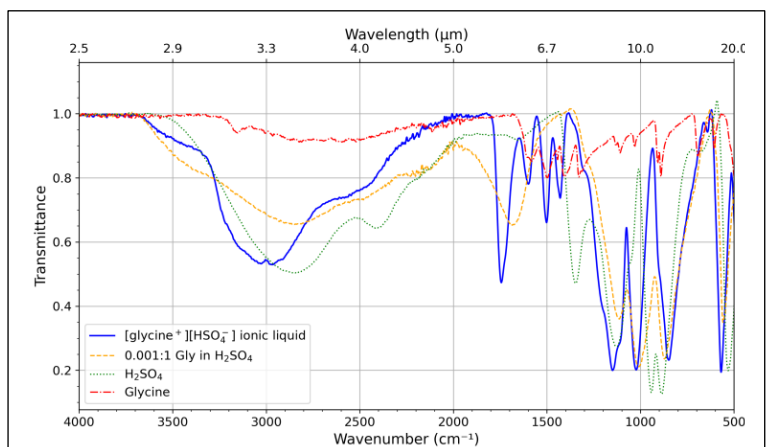

**Figure 3.** FTIR spectra of [glycine$^+$][HSO$_4^-$] ionic liquid and components. The y-axis is transmittance and the x-axis is wavenumber in cm$^{-1}$. The spectrum of [glycine$^+$][HSO$_4^-$] ionic liquid is distinct from the others, namely pure glycine, pure 98% w/w H$_2$SO$_4$ (the rest water), and glycine dissolved in 98% w/w H$_2$SO$_4$, and furthermore is consistent with the spectra of other hydrogen sulfate ionic liquids reported in the literature (e.g., (19)). See also SI Appendix Figures S3 and S4.

FTIR (Figure 3). For all FTIR-measured ionic liquids we detect the $\nu_{S=O}$ and $\nu_{S-O}$ features of $HSO_4^-$ (Table S5). We show that the [glycine$^+$][HSO$_4^-$] ionic liquid differs from Gly dissolved in concentrated sulfuric acid as well as from the individual Gly and H$_2$SO$_4$ components. The pure H$_2$SO$_4$ has a prominent peak at 1345 cm$^{-1}$, corresponding to SO-H bending vibrations. These vibrations are not prominent in the ionic liquid due to the strong ionic bond. Our FTIR spectrum is also consistent with previously reported FTIR spectra of $HSO_4^-$ ionic liquids (19).

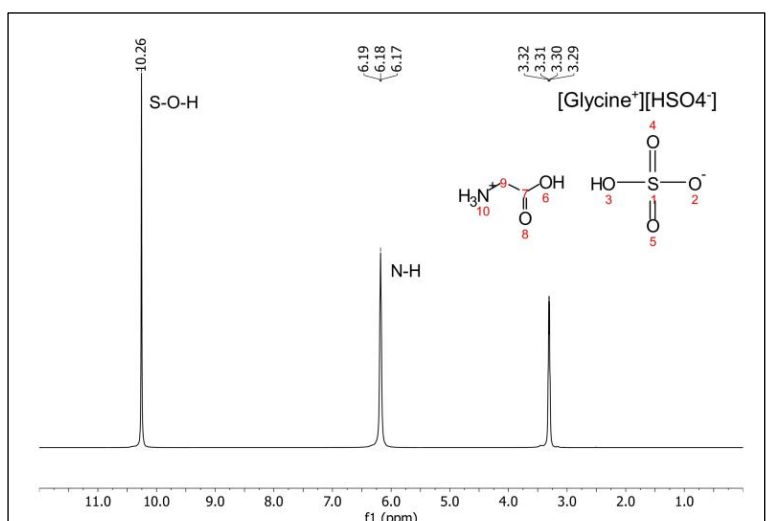

**Figure 4.** $^1H$ NMR spectra of [glycine$^+$][HSO$_4^-$] ionic liquid. The spectra were recorded at 80 °C and 500.18 MHz with DMSO-d$_6$ as the external lock solvent. The peak at 10.26 ppm, corresponds to the SO-H proton of the [HSO$_4^-$] anion while the peak at 6.18 ppm corresponds to the amine hydrogens of the protonated glycine molecule, confirming the identity of the ionic liquid. Additionally, the peak at 3.3 ppm is from the H from the CH$_2$ group of glycine and no other peaks are visible confirms the overall chemical composition of the sample.

Figure S4).

$^1H$ NMR (Figure 4). For all NMR-measured ionic liquids, we identify two sets of protons, in the 6-12 ppm region of the $^1H$ NMR spectrum, that are characteristic of cationic and anionic species. For the [glycine$^+$][HSO$_4^-$] ionic liquid the most downfield-shifted peak, at 10.26 ppm, corresponds to the SO-H proton of [HSO$_4^-$] anion while the peak at 6.18 ppm corresponds to the amine hydrogens of the protonated glycine molecule (SI Section S3 and

Table 1. Thermal decomposition temperatures of selected ionic liquids. $T_0$ is the temperature of onset of decomposition and $T_{50}$ is the temperature at which 50% weight is lost.

| No. | Sample | $T_0$(°C) | $T_{50}$(°C) |
|---|---|---|---|
| 1 | [glycine$^+$][HSO$_4^-$] | 199.44 | 222.66 |
| 2 | [cysteine$^+$][HSO$_4^-$] | 185.91 | 205.98 |
| 3 | [histidine$^+$][HSO$_4^-$] | 180.74 | 281.28 |
| 4 | [valine$^+$][HSO$_4^-$] | 191.4 | 212.58 |
| 5 | [thymine$^+$][HSO$_4^-$] | 248.59 | 299.85 |
| 6 | [guanine$^+$][HSO$_4^-$] | 225.3 | 232.27 |
| 7 | [purine$^+$][HSO$_4^-$] | 229.86 | 297.77 |
| 8 | [diaminopurine$^+$][HSO$_4^-$] | 202.24 | 296.31 |
| 9 | [tetrabutylammonium$^+$][HSO$_4^-$] | 270.78 | 285.99 |
| 10 | [1-ethyl-3-methylimidazolium$^+$][HSO$_4^-$] | 345.65 | 379.91 |
| 11 | [choline$^+$][HSO$_4^-$] | 252.54 | 290.71 |

We also determined the acidity of selected ionic liquids, and find that they are lower than concentrated sulfuric acid but still acidic, with the Gutmann-Beckett acceptor number (AN) values ranging from 117 to 98, as compared to H$_2$SO$_4$ of 122 (SI Section S3 table S6). Our measured AN values are consistent with those reported for other hydrogen sulfate ionic liquids (19). The AN is a quantitative measure of acidity of the solution, determined by the $^{31}$P NMR chemical shift of triethylphosphine oxide (TEPO) when interacting with acid molecules in solution (21, 22). Higher AN values indicate higher acidity of the solution as compared to lower AN values.

We have implicitly described our ionic liquids as equimolar, expected because of the evaporation of excess sulfuric acid under 80−100 °C and low-pressure conditions. To support the equimolar nature of our [glycine$^+$][HSO$_4^-$] ionic liquid we prepared an equimolar ionic liquid directly, by adding equal molar amounts of glycine and H$_2$SO$_4$ (i.e., without the evaporation step) and found a similar AN value (112.8) to the [glycine$^+$][HSO$_4^-$] ionic liquid formed with excess H$_2$SO$_4$ evaporation (113.1).

The ionic liquids have various decomposition temperatures depending on molecules and mixtures, but all higher than about 180–190 °C (See Table 1). Also relevant is that ionic liquids are highly hygroscopic.

## 3. Discussion

### 3.1 Formation of Ionic Liquids in a Planetary Environment

Ionic liquids can conceivably form on a rocky exoplanet when sulfuric acid comes into contact with nitrogen-bearing organic compounds, dissolving them in an environment where excess water and sulfuric acid can evaporate.

**Sulfuric acid.** For hydrogen sulfate ionic liquids to form, a planet must have a source of liquid sulfuric acid. The core production mechanism is the general reaction $SO_2 + H_2O \xrightarrow{[O]} H_2SO_4$ via oxidation—a universal pathway likely to occur on any planet with the necessary reactants. On volcanically active planets, SO$_2$ and H$_2$O vapor can be exsolved from the mantle. The critical step in H$_2$SO$_4$ formation is the oxidation of SO$_2$ to SO$_3$, which requires reactive oxygen species (O) such as O, $O_2$, $O_3$, or $H_2O_2$ to overcome the high activation energy barrier (23). Industrial production of sulfuric acid similarly relies on reactive oxygen and a catalyst to drive this step. On planets, the required oxygen can be supplied by UV photodissociation of oxygen-bearing molecules like $CO_2$ (also volcanically exsolved), $H_2O$, and others. For example, on Venus, UV-driven oxidation of volcanic $SO_2$ to $SO_3$, followed by reaction with trace atmospheric water, produces the planet's dense sulfuric acid clouds (24). An analogous process occurs in Earth's stratosphere, where photochemically generated OH oxidizes $SO_2$ to form $H_2SO_4$ (25). Sulfuric acid may also form at the surface when $SO_2$ reacts with basaltic or mafic rocks. On Earth, $Fe_2O_3$-bearing rocks catalyze this reaction in $SO_2$-rich volcanic environments, such as at Kīlauea, Hawai'i (26). Atmospheric $O_2$ likely replenishes lattice oxygen in the $Fe_2O_3$ catalyst; a similar mechanism could operate on exoplanets without $O_2$, though this remains to be investigated (e.g., volcanic $NO_2$, such that $SO_2 + NO_2 \rightarrow SO_3 + NO$ for an overall $SO_2 + 2NO_2 + H_2O \rightarrow H_2SO_4 + 2NO$). Sulfuric acid has also been observed on Europa, where it forms through radiolytic oxidation of surface sulfur (reviewed by (23)). In summary, volcanically exsolved $SO_2$ and $H_2O$ can generate sulfuric acid through diverse pathways, and concentrated $H_2SO_4$ can condense even under low atmospheric pressure.

The availability of hydrogen may initially seem problematic on a water-depleted planet. However, Venus serves as a counterexample: despite limited hydrogen, it hosts abundant sulfuric acid. On Venus, both $SO_2$ and $H_2O$ are outgassed from volcanoes, and UV radiation drives their conversion to $H_2SO_4$. The $H_2O/SO_2$ ratio in Venusian volcanic gases is relevant but not well constrained. Zolotov et al. (27) report that $SO_2$ dominates the molar fraction over water, estimating $SO_2$ could approach 10%, with $CO_2$ making up the remaining 90%. They note that water is not considered an important volcanic gas on Venus. Modeling by Constantinou et al. (28) places an upper limit of 6% for the water mole fraction. By contrast, Earth's volcanic emissions are ~90% water by mole fraction, making $H_2O$ the dominant gas (29). These comparisons underscore that sulfuric acid can plausibly form even in hydrogen-limited environments. The work of Janssen et al. (30) supports the argument that $H_2SO_4$ is likely to exist in sulfur-rich, oxygen-rich, and hydrogen-poor planetary conditions. See Table S7 in the SI.

$H_2SO_4$ is a liquid over a wide range of temperature and pressure conditions. The key question for ionic liquid formation is whether liquid sulfuric acid can exist at the planetary surface long enough to dissolve organic compounds, then evaporate efficiently to leave behind an ionic liquid. The answer is yes—formation and evaporation is viable across a broad range: from low pressures and temperatures (~300 K at $10^{-6}$ atm) to higher pressures and temperatures (~360–470 K at $10^{-2}$ atm, pressures comparable to Mars-like conditions). We adopt 470 K as an upper temperature cutoff, not because sulfuric acid cannot exist as a liquid above this temperature, but because many of the ionic liquids we study begin to degrade beyond that point (see Table 1). A phase diagram for sulfuric acid with estimated boiling curves is shown in Figure 5, adapted from (31) and supplemented by our own estimated boiling points from the vapor pressure–temperature relationship via the Antoine equation and data from (32, 33). For example, we estimate that concentrated sulfuric acid can evaporate within a month (a subjective choice) within the range of about 100 K cooler than the boiling point, based on our experimental evaporation measurements (Table S9).

Although we favor the warm, thin atmosphere exoplanet as a host to hydrogen sulfate ionic liquids, our extended experiments on basalt rocks (Figure S19) show that even at room temperature, room pressure, room humidity conditions ionic liquid can form once the sulfuric acid dissolves the N-containing organic—as long as the excess sulfuric acid disperses by evaporation, spreading into rock pores, reaction with the rock, etc.

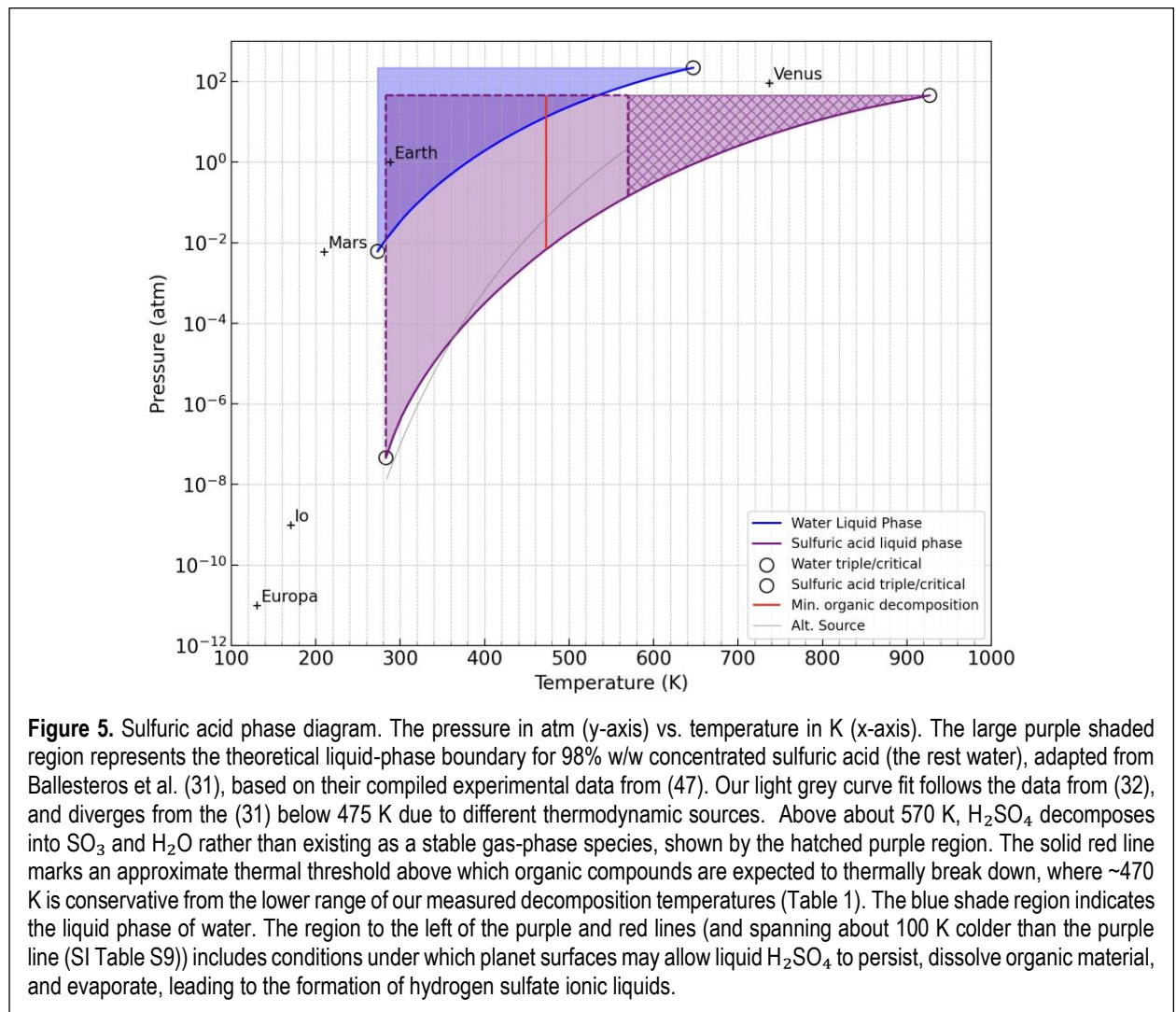

**Figure 5.** Sulfuric acid phase diagram. The pressure in atm (y-axis) vs. temperature in K (x-axis). The large purple shaded region represents the theoretical liquid-phase boundary for 98% w/w concentrated sulfuric acid (the rest water), adapted from Ballesteros et al. (31), based on their compiled experimental data from (47). Our light grey curve fit follows the data from (32), and diverges from the (31) below 475 K due to different thermodynamic sources. Above about 570 K, $H_2SO_4$ decomposes into $SO_3$ and $H_2O$ rather than existing as a stable gas-phase species, shown by the hatched purple region. The solid red line marks an approximate thermal threshold above which organic compounds are expected to thermally break down, where ~470 K is conservative from the lower range of our measured decomposition temperatures (Table 1). The blue shade region indicates the liquid phase of water. The region to the left of the purple and red lines (and spanning about 100 K colder than the purple line (SI Table S9)) includes conditions under which planet surfaces may allow liquid $H_2SO_4$ to persist, dissolve organic material, and evaporate, leading to the formation of hydrogen sulfate ionic liquids.

**Surface organics.** For an ionic liquid to form, concentrated sulfuric acid must dissolve nitrogen-containing organics. The planet must therefore have pockets of organic material on its surface. So far, organics have been detected on every solid-surfaced body where they have been specifically searched for (Mercury, the Moon, Mars, Ceres, Ganymede and Callisto, Enceladus, Titan, Pluto and Charon, comets, and asteroids). How these organics formed—and their chemical character—remains an outstanding question in planetary science. For details and relevance to ionic liquid formation, see Table S10.

On the vast majority of these bodies, organics' concentrations are not measured; however, detection itself implies substantial local deposits. For example, on Mercury, a ~10 cm deep layer of organics covers the water ice deposits in the polar craters (34). Comet 67P has an organic carbon content of ~45 wt% in dust particles ranging from 50–1000 μm in size (35), showing that at the microscale, organic matter can be highly concentrated. While the spatial extent and exact composition of organics on these cometary particles remains unmeasured (e.g., surface coverage, clumping, N-containing percentage, etc.), their presence confirms the existence of dense organic deposits. On meteorites, organic matter can occur in globular deposits with variable sizes, typically ranging from 50–300 nm, and is often highly concentrated (36).

In our own solar sdropletystem no bodies are known to have ionic liquids. Io does not have liquid sulfuric acid, likely both because it is severely H-depleted and too cold for liquid sulfuric acid. Mercury does not have any active volcanoes so current formation of $H_2SO_4$ is not possible. Notably, Europa has solid sulfuric acid on its surface, showing that $H_2SO_4$ does not sublimate even at $10^{-12}$ atm under frigid temperatures.

### 3.2 Atmosphereless Worlds

The range of exoplanet surface conditions may extend down to atmosphereless worlds, enabled by the extremely low vapor pressure of ionic liquids. Even a small droplet of many ionic liquids will not "dry out," even under vacuum at Earth-like temperatures (37). This makes the persistence of ionic liquids plausible on surfaces lacking atmospheric pressure. However, exposure to the harsh space environment could lead to space weathering of the surface material. Phenomena such as micrometeorite impacts, solar wind plasma, UV and high energy radiation could alter the chemical and physical properties of the unprotected planetary surface, including any surface ionic liquid (38, 39). The penetration depth of space weathering ranges from 2−3 topmost atomic layers, due to ion sputtering, to 100 nm for extremely high energy protons. On the other hand, the depth of penetration of galactic cosmic rays with GeV energy is much larger than high energy protons and can reach 1 meter (38). Thus, any sub-surface ionic liquid, even in shallow depths, could be protected from most harsh space weathering processes. A protective magnetic field could help shield the surface, and additional protection could come from subsurface environments, shaded regions, or crevices where ionic liquids can be shielded from direct exposure. For example, ionic liquids might survive on the dark side of a tidally locked planet near warm volcanic vents, assuming the planet maintains a magnetic field. The long-term survival of ionic liquids—particularly their resistance to UV and energetic particle bombardment—is an open question and a topic for future study.

### 3.3 A Range of Ionic Liquids

Hydrogen sulfate ionic liquids might just be one example out of many possible planetary ionic liquids. In principle any positively charged organic molecule could act as a cation in an ionic liquid pair. There are also at least several possible anions that could pair with organic cations, such as $Cl^-$, $NO_3^-$, $ClO_4^-$, $HCOO^-$, $CH_3COO^-$, and deep eutectic solvents which are related to ionic liquids but do not have charged components. Indeed, none of the planetary bodies in our own solar system are expected to have surface ionic liquids because they are either too cold for the liquid phase or lack the necessary chemical components.

## 3.4 Habitability

We now turn to support our title statement "habitable world". The building blocks of our water-based biochemistry, such as DNA and proteins (enzymes) are often stable and functional in the presence of ionic liquids (reviewed in (40, 41)), and some even function in dehydrated pure ionic liquids (e.g., reviewed in (42))). Finally, many ionic liquids are not toxic (43) and are chemically benign and unreactive when in contact with Earth life biochemicals (e.g., (42)).

Ionic liquid cationic components like quaternary amines and other charged N-containing organic compounds are readily made by life and the anionic components of ionic liquids, such as $H_2SO_4$, can be present in the right planetary environment. Ionic liquids are polar and can dissolve salts and complex polymers, which meets some of the criteria suitable for a solvent for life (13). Many species use or produce compounds that could be used to make ionic liquids, such as amines or sugars, as a means to survive desiccation (e.g., (44)). Moreover, because ionic liquids do not evaporate the risk of permanent desiccation for an organism using an ionic liquid as a solvent does not exist, leaving only the risk of mechanical damage as a cause of solvent loss to the outside world.

A planet may initially have only small isolated pockets of ionic liquid, due to the limiting amount of input ingredients, sulfuric acid and organics. Therefore, the ionic liquid is likely itself not a dominant planetary surface feature, meaning it is likely currently undetectable. If there are large reservoirs of surface ionic liquid, future telescopes that can spatially resolve the surface during secondary eclipse ingress and egress may be able to detect the sulfate group via the planet's emission spectra. If life permeates the surface, perhaps its pigments would show a feature in reflected light (e.g., vegetation's "red edge"). Life itself might generate waste gases from exploiting chemical redox gradients in the environment. Any gas would diffuse upwards and get photodissociated, with atoms escaping the planet. It is possible that a powerful UV telescope is able to spot such an exosphere with unusual ratios of atoms that could be further explored as a sign of life.

In summary, ionic liquids can exist as stable liquids under surface conditions that are too warm or too low in pressure for liquid water to persist. This includes warm planets with thin atmospheres and possibly even atmosphereless bodies, where the extremely low vapor pressure of ionic liquids prevents them from evaporating if protected from UV and harsh cosmic radiation. Liquid is a fundamental requirement for life (45), and the ability of ionic liquids to form and remain stable in these environments expands the range of planetary conditions where life could potentially arise. Ionic liquid formation from volcanic sulfuric acid and surface organics—both observed in the laboratory and now expected on some rocky bodies—introduces a new class of potentially liquid-bearing planets. This significantly broadens the concept of planetary habitability beyond the traditional water-based paradigm.

## 4. Materials and Methods

### 4.1 Materials

We purchased sulfuric acid (95-98%), glycine (Cat. No. G7126), histidine (Cat. No. H8000), thymine (Cat. No. T0376), adenine (Cat. No. A8626), purine (Cat. No. P55805), guanine (Cat. No. G11950), diaminopurine (Cat. No. 247847), tetrabutylammonium hydroxide (Cat. No. 178780), acetic acid (Cat. No. 695092), oxalic acid (Cat. No. 241172), benzoic acid (Cat. No. 242381), succinic acid (Cat. No. S9512), stearic acid (Cat. No. S4751), palmitic acid (Cat. No. P0500), and dimethyl sulfoxide-$d_6$ (1034240100) from Sigma-Aldrich. The remaining amino acids—alanine, arginine, asparagine, aspartic acid, cystine, lysine, methionine, tryptophan, valine, glutamine, glutamic acid, isoleucine, leucine, phenylalanine—were part of a kit of 20 amino acids from Sigma-Aldrich (Cat. No. LAA21-1KT). Choline-hydroxide (Cat. No. 224210) was purchased from Beantown Chemicals and 1-Ethyl-3-methylimidazolium hydrogen sulfate (H27232) and pyrimidine (157700050) from Thermo Scientific Chemicals. The triethylphosphine oxide (Cat. No. AA3039103) was purchased from Fisher Scientific. Alanylglycine dipeptide (Cat. No. ALA121374-5G) and basalt rock samples Ward's Basalt (Cat. No. 470025-914) were purchased from VWR.

**4.2 Preparation of Ionic Liquids:** To form equimolar ionic liquids, we mixed approximately equal moles of the organic compound and 98% w/w concentrated sulfuric acid (the rest water). We then added about 5% extra $H_2SO_4$ to ensure the organic completely dissolved in sulfuric acid. To further aid solubility, the mixture was incubated in a sand bath at 80+/-5 °C for 12−24 hours. The time required to dissolve varies depending on the organic molecule. Once the organics are completely dissolved, we evaporated the excess sulfuric acid from the mixture inside a custom-made low-pressure evaporator system (100+/-5 °C and $10^{-5}$ bar pressure). These temperature pressure conditions allow excess sulfuric acid to evaporate slowly. More details on quantities and evaporation time are provided in SI, Section 2. The remaining substance is a highly viscous ionic liquid. We followed the same procedure in preparation of all other ionic liquids from all other nitrogen-containing compounds.

Here we have described our method for forming ionic liquids, we emphasize that the initial concentrations of organics in sulfuric acid, the exact evaporation temperatures and pressures, and the timescales can be widened. What matters is the complete evaporation of excess concentrated sulfuric acid (which depends on temperature and pressure, see SI, Section S2 and Table S9). We emphasize that equimolar ionic liquids can form even at room temperature and pressure, in the situation where there is no excess sulfuric acid to be evaporated. In the lab, we optimize for efficiency by using excess sulfuric acid to aid dissolution of the organic (typically in powder form), and evaporating at our pumping system's minimum pressure, but the underlying conclusion remains: initial concentrations are not critical.

**4.3 $^1$H NMR Spectroscopy:** We used $^1$H NMR spectra to confirm the structure of selected ionic liquids. The ionic liquids were transferred to NMR tubes (5 mm, borosilicate glass). A sealed capillary tube containing a deuterated solvent dimethyl sulfoxide-$d_6$ (DMSO-$d_6$), an external lock solvent, was inserted into the NMR tube containing the ionic liquid. We obtained the $^1$H NMR spectra at 80 °C using a Bruker Avance Neo spectrometer operating at 500.18 MHz.

**4.4 Acceptor Number Determination:** We used the Gutmann Beckett Acceptor Number method to determine the Lewis acidity of selected ionic liquids (21, 22). We used triethylphosphine oxide (TEPO) as the probe for $^{31}$P NMR. We added approximately 50 mg of TEPO to ~1 ml of the ionic liquid and incubate the sample at 80 °C to allow mixing. The samples were transferred into 5 mm NMR tubes with capillary tube insert containing DMSO-$d_6$ as external lock solvent. $^{31}$P NMR spectra were obtained at 80 ℃ using a Bruker Avance Neo spectrometer operating at 500.18 MHz. We measured the spectra for $H_2O$ and $H_2SO_4$ to establish the range of acidity within which the ionic liquid acidity must lie.

**4.5 FTIR Spectroscopy:** Infrared spectra of ionic liquids, the organics, and $H_2SO_4$ were measured using the Bruker Alpha II FTIR spectrometer with a Diamond Crystal ATR (Attenuated Total internal Reflectance) accessory. We acquired 64 scans for each sample with 2 cm$^{-1}$ resolution. See SI, Section 2 for details.

**4.6 Thermogravimetric Analysis (TGA):** We used the TGA method to determine the thermal decomposition onset temperature of selected ionic liquids. About 1−2 mg of ionic liquid was loaded onto a platinum pan. The samples were heated from 30 °C to 600 °C with a rate of 10 °C/min under the flow of nitrogen. The weight change was measured using TA Instruments Discovery TGA 5500.

**Data, Materials, and Software Availability.** The original data files have been deposited in Zenodo: https://zenodo.org/records/15596466.

**Acknowledgements**

This research benefitted from the use of TGA TA 5500 made available by the Institute for Soldier Nanotechnologies, a U.S. Army-sponsored University Affiliated Research Centers (UARC) at MIT. This research benefited from the use of FTIR and NMR spectroscopy instruments at the Massachusetts Institute of Technology (MIT) Department of Chemistry Instrumentation Facility and we thank the Director Walter Massefski and staff Sarah Willis and Bridget Becker. We thank the four reviewers for helpful comments which improved the paper. This work was partially funded by The Alfred P. Sloan Foundation grant G-2023-20929, The Volkswagen Foundation grant 9E126, and Nanoplanet Consulting.

**Supplementary Information for**

Warm, Water-Depleted Rocky Exoplanets with Surface Ionic Liquids: A Proposed Class for Planetary Habitability

Rachana Agrawal[1], Sara Seager[1,2,3,4,*], Iaroslav Iakubivskyi[1,5], Weston P. Buchanan[1], Ana Glidden[1,4], Maxwell D. Seager[6], William Bains[7], Jingcheng Huang[1], Janusz J. Petkowski[8,9,*]

[1] Department of Earth, Atmospheric and Planetary Sciences, Massachusetts Institute of Technology, 77 Massachusetts Avenue, Cambridge, MA 02139, USA
[2] Department of Physics, Massachusetts Institute of Technology, 77 Massachusetts Avenue, Cambridge, MA 02139, USA
[3] Department of Aeronautical and Astronautical Engineering, Massachusetts Institute of Technology, 77 Massachusetts Avenue, Cambridge, MA 02139, USA
[4] Kavli Institute for Astrophysics and Space Research, Massachusetts Institute of Technology, 77 Massachusetts Avenue, Cambridge, MA 02139, USA
[5] Tartu Observatory, University of Tartu, 1 Observatooriumi, 61602 Tõravere, Estonia
[6] Department of Chemistry and Biochemistry, Worcester Polytechnic Institute, Worcester, MA 01609, USA
[7] School of Physics and Astronomy, Cardiff University, 4 The Parade, Cardiff CF24 3AA, UK
[8] Faculty of Environmental Engineering, Wroclaw University of Science and Technology, 50-370 Wroclaw, Poland
[9] JJ Scientific, Warsaw, Mazowieckie, Poland

*Corresponding author: Sara Seager, Janusz J. Petkowski
**Email:** seager@mit.edu, janusz.petkowski@pwr.edu.pl

**This PDF file includes:**

Supplementary text
Figures S1 to S19
Tables S1 to S10
Supplementary References

# Supplementary Information Text

## S1. Ionic Liquids Summary

Ionic liquids are formally defined as salts with melting temperatures below 100 °C (1). Ionic liquids are synthetic and have not previously been considered as naturally occurring substances. We show that ionic liquids form with planetary materials: a wide range of N-containing organics as the cation, paired with a hydrogen sulfate anion (hydrogen sulfate salts). Here we summarize all the organics and mixtures we tried (Table S1). The 20 biogenic amino acids form ionic liquids at room temperature. The nucleic acid bases we tried also form hydrogen sulfate salts but were solid at room temperature; a subset were liquid at 100 °C, while others remained solid. We also show that mixtures of two or more organics with at least one N-containing organic also form ionic liquids.

Additionally, we tested six carboxylic acids (aliphatic and aromatic) as cations for ionic liquids. We hypothesized that molecules with a carbonyl group might also form ionic liquids with concentrated sulfuric acid (2), as the oxygen will gain a positive charge due to the protonation of the carbonyl oxygen (3–5). We found that they do not form ionic liquids under tested conditions.

We further characterize selected ionic liquids using a variety of analytical chemical techniques. We determine the composition, decomposition temperatures, and acidity by using $^1$H Nuclear Magnetic Resonance (NMR) spectroscopy, Fourier transform infrared (FTIR) spectroscopy, Thermogravimetric Analysis (TGA), and $^{31}$P NMR, respectively. More details of the experiments and analysis results are given in subsequent sections. Table S1 points towards the relevant sections for the selected ionic liquids.

To evaluate the plausibility of ionic liquid formation under diverse planetary conditions, we conducted a set of experiments examining chemical composition, environmental variability, and physical constraints. We varied the concentrations of both sulfuric acid and organics, assuming organics are the limiting component. We tested a range of organic concentrations diluted in concentrated sulfuric acid (Table S2), including as low as 20 mM glycine, which produced a similar FTIR spectrum to the equimolar mixture. Given the hygroscopic nature of ionic liquids, we also tested sulfuric acid diluted in water (0–95% w/w $H_2SO_4$, the rest water), and observed that, after evaporation of excess water and sulfuric acid, the resulting ionic liquid was qualitatively similar across this concentration range (Section S4). To examine whether ionic liquids can form at planetary-relevant microscale levels, we attempted to generate the smallest detectable quantities, constrained only by FTIR detection limits (Section S4). In addition, we expanded tests on rock surfaces (beyond Figure 2 and related discussion) to assess formation in more complex, non-pure environments (Section S3). While we conservatively set an upper temperature limit of 200 °C based on observed decomposition of the most labile compounds, this threshold likely varies; more thermally stable ionic liquids may persist at higher temperatures. Overall, our results highlight the robustness of ionic liquid formation across a wide range of starting conditions, supporting their plausibility on warm, water-depleted planetary surfaces.

Table S1. Summary of the hydrogen sulfate (HSO$_4^-$) ionic liquid formation for selected organic compounds. See Section S3 for more details on the characterization and analysis of selected ionic liquids. RT is room temperature.

| No. | Organic Components | Ionic Liquid | Visual Observation | Analysis |
|---|---|---|---|---|
| **Amino Acids** | | | | |
| 1 | Alanine | Yes | Clear, viscous | $^1$H NMR, FTIR |
| 2 | Arginine | Yes | Clear, viscous | Visual |
| 3 | Asparagine | Yes | Clear, viscous | Visual |
| 4 | Aspartic acid | Yes | Yellowish, viscous | Visual |
| 5 | Cysteine | Yes | Clear, viscous | $^1$H NMR, FTIR, TGA |
| 6 | Glutamine | Yes | Clear, viscous | Visual |
| 7 | Glutamic acid | Yes | Clear, viscous | Visual |
| 8 | Glycine | Yes | Clear, viscous | $^1$H NMR, FTIR, TGA, $^{31}$P NMR |
| 9 | Histidine | Yes | Yellowish, highly viscous | FTIR, TGA |
| 10 | Isoleucine | Yes | Clear, viscous | Visual |
| 11 | Leucine | Yes | Clear, viscous | Visual |
| 12 | Lysine | Yes | Clear, viscous | Visual |
| 13 | Methionine | Yes | Red, viscous | Visual |
| 14 | Phenylalanine | Yes | Clear, viscous | Visual |
| 15 | Proline | Yes | Clear, viscous | Visual |
| 16 | Serine | Yes | Clear, viscous | Visual |
| 17 | Threonine | Yes | Clear, viscous | Visual |
| 18 | Tryptophan products | Yes | Dark, viscous | $^1$H NMR |
| 19 | Tyrosine | Yes | Yellowish, viscous | Visual |
| 20 | Valine | Yes | Clear, viscous | $^1$H NMR, FTIR, TGA |
| **Nucleic Acid Bases** | | | | |
| 21 | Purine | Yes | Solid at RT, liquid at 100 °C | FTIR, TGA |
| 22 | Adenine | Yes | Solid at RT, liquid at 100 °C | Visual |
| 23 | Guanine | Yes | Solid at RT, liquid at 100 °C | FTIR, TGA |
| 24 | Diaminopurine | No | Solid at 100 °C | TGA |
| 25 | Pyrimidine | No | Solid at 100 °C | Visual |
| 26 | Thymine | Yes | Solid at RT, liquid at 100 °C | $^1$H NMR, FTIR, TGA |
| **Carboxylic Acids** | | | | |
| 27 | Acetic acid | No | Completely evaporates | Visual |

| 28 | Benzoic acid | No | Completely evaporates | Visual |
|---|---|---|---|---|
| 29 | Oxalic acid | No | Completely evaporates | Visual |
| 30 | Succinic acid | No | Completely evaporates | Visual |
| 31 | Stearic acid | No | Completely evaporates | Visual |
| 32 | Palmitic acid | No | Completely evaporates | Visual |
| **Other N-containing Organics** | | | | |
| 33 | Tetrabutylammonium | Yes | Clear | $^1$H NMR, FTIR, TGA, $^{31}$P NMR |
| 33 | 1-ethyl-3-methylimidazolium | Yes | Clear, gel-like | $^1$H NMR, FTIR, TGA |
| 33 | Choline | Yes | Brown, with solid precipitates | FTIR, TGA |
| 34 | Alanylglycine | Yes | Clear, viscous | $^1$H NMR, FTIR |
| **Mixtures** | | | | |
| 35 | Glycine, thymine | Yes | Clear, viscous | $^1$H NMR, FTIR |
| 36 | Glycine, acetic acid | Yes | Acetic acid evaporates | FTIR |
| 37 | Glycine, saccharose | Yes | Dark, Saccharose decomposes | FTIR |
| 38 | Glycine, thymine, stearic acid, paraffin, naphthalene | Yes | Dark, semi-solid | FTIR |

Table S2. Summary of experiments to test the formation of ionic liquid in various starting conditions.

| No. | Experiment | Parameters | Analysis |
|---|---|---|---|
| 1 | Varying glycine concentration in concentrated sulfuric acid | Gly:$H_2SO_4$ molar ratio of 1:1, 1:2, 1:3, 0.001:1 | Sec. S4, Fig. S14, S15, S16 |
| 2 | Varying $H_2SO_4$ concentration in water | wt% of $H_2SO_4$ in $H_2O$ 0%, 5%, 25%, 50%, 75%, 95% | Sec. S4, Fig S13 |
| 3 | Microscale formation of [glycine$^+$][$HSO_4^-$] ionic liquid | 300 µg of glycine in 20 µL of 98% w/w $H_2SO_4$, the rest water | Sec. S4, Fig S16 |
| 4 | [glycine$^+$][$HSO_4^-$] formation on basalt | Varying concentration of glycine and $H_2SO_4$; storage in the open, a desiccated container, and evaporator conditions | Sec. S4, Fig S17, S18 |

## S2. Ionic Liquid Formation Methods

We describe the materials used, detailed experimental setup and procedure. See also Methods Section in the main manuscript.

### Materials

This paragraph is repeated from the main manuscript for completeness. We purchased sulfuric acid (95-98%), glycine (Cat. No. G7126), histidine (Cat. No. H8000), thymine (Cat. No. T0376), adenine (Cat. No. A8626), purine (Cat. No. P55805), guanine (Cat. No. G11950), diaminopurine (Cat. No. 247847), tetrabutylammonium hydroxide (Cat. No. 178780), acetic acid (Cat. No. 695092), oxalic acid (Cat. No. 241172), benzoic acid (Cat. No. 242381), succinic acid (Cat. No. S9512), stearic acid (Cat. No. S4751), palmitic acid (Cat. No. P0500), and dimethyl sulfoxide-$d_6$ (1034240100) from Sigma-Aldrich. The remaining biogenic amino acids—alanine, arginine, asparagine, aspartic acid, cystine, lysine, methionine, tryptophan, valine, glutamine, glutamic acid, isoleucine, leucine, phenylalanine—were part of a kit of 20 amino acids from SigmaAldrich (Cat. No. LAA21-1KT). Choline-hydroxide (Cat. No. 224210) was purchased from Beantown Chemicals and 1-Ethyl-3-methylimidazolium hydrogen sulfate (H27232) and pyrimidine (157700050) from Thermo Scientific Chemicals. The triethylphosphine oxide (Cat. No. AA3039103) was purchased from Fisher Scientific. Alanylglycine dipeptide (Cat. No. ALA121374-5G) and basalt rock samples Ward's Basalt (Cat. No. 470025-914) are purchased from VWR.

**The Thermo-vacuum Evaporation Setup**

The custom evaporator (placed inside the fume hood) used for the removal of excess sulfuric acid under low pressure is shown in Figure S1. The setup includes the following main components. (1) A rotary vane vacuum pump prepared to be used with corrosion-resistant PFPE oil (Alcatel 2010SD Mechanical Pump). The pump is capable of a base pressure of 0.01 mbar. (2) The vacuum chamber consists of a borosilicate glass jar (VWR Cat. No. 89055-058), resistant to sulfuric acid, and allows observation of the evaporation process. (3) A hot plate with temperature control (VWR Cat. No. 76447-036). We experimentally determined the evaporation rate of 80%, 90%, and 98% w/w sulfuric acid, the rest water (without any dissolved solute) in the custom evaporator (Figure S1). The temperature was varied from 90-125 °C in 5 °C intervals and three data points were obtained at each temperature and concentration condition. For every data point, 20 µL sulfuric acid was pipetted onto a glass dish and the area over which it spread was measured. The pressure reading and time required for complete evaporation of sulfuric acid was calculated. We used this data to further calculate the minimum time required to evaporate sulfuric acid from a mixture of organic and sulfuric acid.

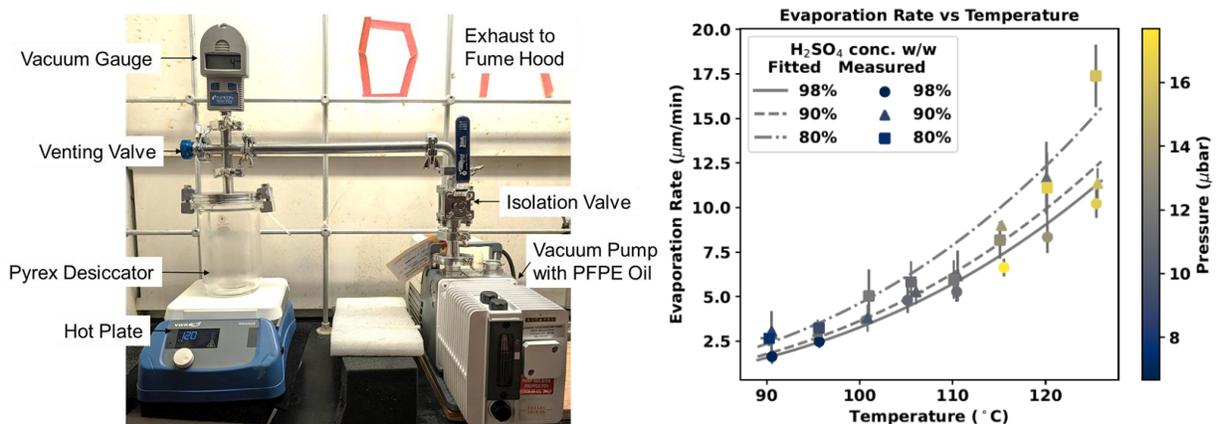

Figure S1. Left panel: Thermo-vacuum evaporation setup with the key components labeled. Right panel: Evaporation rate of sulfuric acid (without any solute) in µm/min (or $10^4 \times mL/cm^2/min$) on the y-axis as a function of temperature in °C on the x-axis. The colorbar on the right indicates the average pressure in µbar inside the chamber for each measurement.

**Ionic Liquid Production Process**

In this section we summarize the procedure for ionic liquid production. We describe the formation of hydrogen sulfate ionic liquids in equimolar (1:1 molar, also called neat), as well as the non-equimolar variants.

To form equimolar ionic liquids, we mixed approximately equal molar amounts of the organic compound and 98% w/w sulfuric acid (the rest water). We then added about 5% extra pure $H_2SO_4$ to ensure the complete dissolution of organic molecules in concentrated sulfuric acid. For example, to form an equimolar [glycine$^+$][HSO$_4^-$] ionic liquid, we mixed 1.4 g of glycine into 1.05 mL of pure 98% w/w $H_2SO_4$ (the rest water). To ensure complete dissolution of organic compounds in concentrated sulfuric acid we incubated the mixture in a sand bath at 80 °C +/-5 °C for 12-24 hours. This initial incubation time does not affect the ionic liquid formation experiment. Once the organics were completely dissolved, we evaporated the excess sulfuric acid from the mixture inside the custom thermo-vacuum evaporator setup (100 °C +/-5 °C and 0.01 mbar pressure). We chose these temperature-pressure conditions to allow for slow evaporation of excess sulfuric acid. We left the evaporation system running overnight. The exact time did not matter as long as it was long enough to evaporate the excess sulfuric acid. Once the ionic liquid formed, the ionic liquid did not evaporate. To estimate the minimum time to ensure complete evaporation of excess sulfuric acid, we used the evaporation rate data for concentrated sulfuric acid alone (Figure S1) and ran the ionic liquid formation experiment for at least twice the required calculated time. For example, to form an equimolar [glycine$^+$][HSO$_4^-$] ionic liquid, the extra 50 µL of $H_2SO_4$, beyond the 1:1 molar ratio, requires 2 hours to evaporate. Therefore, we ran the evaporator overnight (>12 hours) to ensure complete evaporation of excess sulfuric acid. We confirmed that the ionic liquid does not evaporate by leaving a ~10 µL glycine- $H_2SO_4$ sample in the evaporator (100 °C +/-5 °C and 0.01 mbar pressure) for 24 hours and it does not evaporate.

We follow the same procedure in preparation of all the ionic liquids from all other organic compounds listed in Table S1.

The exact temperature-pressure conditions for preparation of our ionic liquids do not matter as long as the temperature is below the decomposition temperatures of organic compounds (Table 1 in the main manuscript) and the pressure is low enough to ensure evaporation of excess sulfuric acid. See Section S4 for further data on the ionic liquid formation procedures and the estimation of the evaporation rate of concentrated sulfuric acid.

For the non-equimolar ionic liquid mixtures (1:2, 1:3) and the diluted glycine sample (0.001:1) we followed the same procedure as for the 1:1 equimolar ionic liquid formation. We adjusted the required volume of liquid concentrated sulfuric acid accordingly (Table S3). See Section S4 for further data on the ionic liquid formation procedures and the estimation of the evaporation rate of concentrated sulfuric acid.

Table S3. Amounts of glycine and concentrated sulfuric acid (98% w/w $H_2SO_4$, the rest water) used in ionic liquid formation.

| Molar Ratio | Glycine (mg) (conc. in M) | $H_2SO_4$ (mL) Before Evaporation |
|---|---|---|
| 1:1 | 1400 (~18M) | 1.05 |
| 0.001:1 | 1.4 (20mM) | 1 |
| 1:2 | 1400 (~9M) | 2 |
| 1:3 | 1400 (~6M) | 3 |

## S3. Ionic Liquid Chemical Analysis Methods and Results

In this section we provide further details on ionic liquid chemical characterization and analysis. We describe the methods and additional supplementary results for $^1$H NMR, FTIR, TGA and acidity assessment ($^{31}$P NMR) for a set of selected ionic liquids (Table S1 and Figure S2).

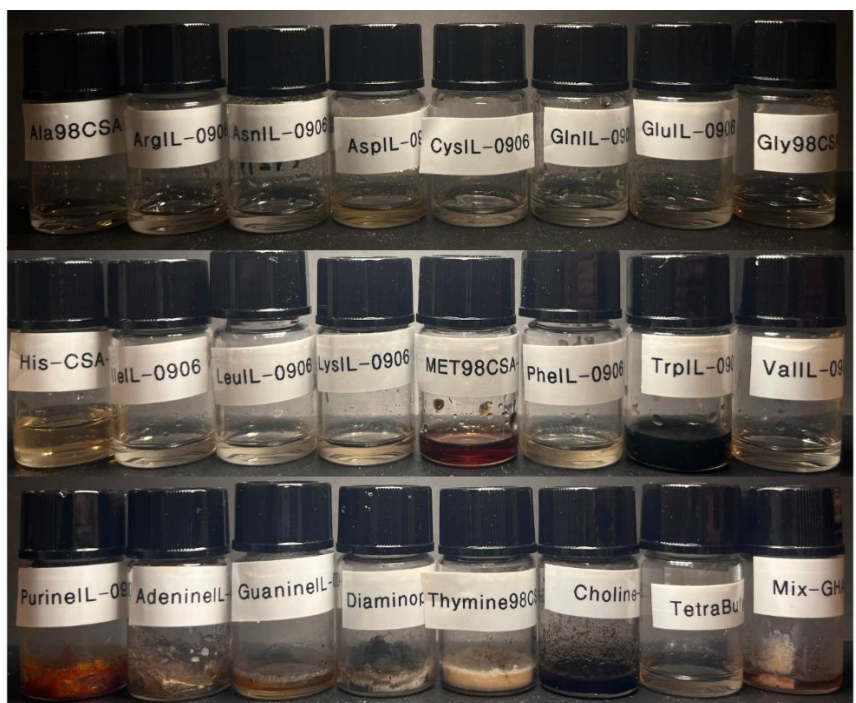

Figure S2. Vials of evaporation products composed of N-containing organics and hydrogen sulfate ($HSO_4^-$) anions. The starting organic compounds are labeled on each vial. The ionic liquids were generated by evaporating excess sulfuric acid from mixtures of various organic compounds and 98% w/w sulfuric acid (the rest water). Note that even compounds that are unstable in concentrated sulfuric acid, like tryptophan (6), that react to form a highly complex mixture of diverse organics, do form ionic liquids upon evaporation of excess sulfuric acid.

# $^1$H Nuclear Magnetic Resonance (NMR) Spectroscopy

This paragraph is repeated from the main manuscript and expanded upon. We use $^1$H NMR spectra to further support the characterization of the selected ionic liquids. The neat ionic liquids were transferred to NMR tubes (5 mm, borosilicate glass). Approximately 300-400 µL ionic liquid was pipetted into the NMR tubes. A sealed capillary tube containing a deuterated solvent dimethyl sulfoxide-$d_6$ (DMSO-$d_6$), as an external lock, was inserted into the NMR tube containing the ionic liquid. We obtained the $^1$H NMR spectra at 80 °C using a Bruker Avance Neo spectrometer operating at 500.18 MHz (Figure S3 and Figure S4).

We assigned the peaks corresponding to SO-H and N-H for each tested ionic liquid (Table S4). In all the spectra, the SO-H peak is the most downfield shifted. For all NMR-measured ionic liquids (with the exception of tetrabutylammonium and tryptophan ionic liquid samples) we identify two sets of protons in the 6-12 ppm region of the $^1$H NMR spectrum that are characteristic of cationic and anionic species. For the [glycine$^+$][HSO$_4^-$] ionic liquid the most downfield-shifted peak, at 10.26 ppm, corresponds to the SO-H proton of [HSO$_4^-$] anion while the peak at 6.18 ppm corresponds to the amine hydrogens of the protonated glycine molecule. We note that the tetrabutylammonium$^+$ [TBA$^+$] cation does not have a N-H group; thus, no N-H peak is visible.

We included the tryptophan-based ionic liquid to show that even in a complex mixture of many N-containing organic compounds, an ionic liquid will still form upon evaporation of excess sulfuric acid (Figure S2 and Figure S3). Tryptophan is unstable and reacts readily with concentrated sulfuric acid (6). The ionic liquid that forms from tryptophan and sulfuric acid is a product of that reactivity. The resulting $^1$H NMR therefore is very "crowded" as it contains many different reaction products. The identity of the positively charged organic molecule that forms the ionic liquid is unknown, but it has to be one, or more, product of tryptophan decomposition. The identification of these products is beyond the scope of this work.

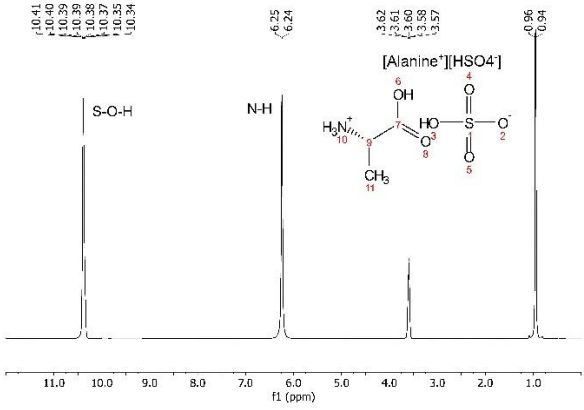
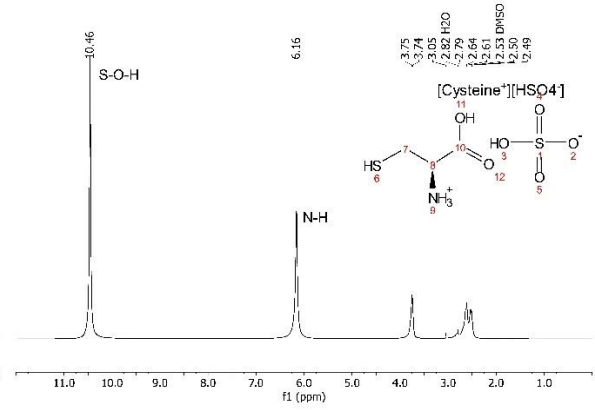
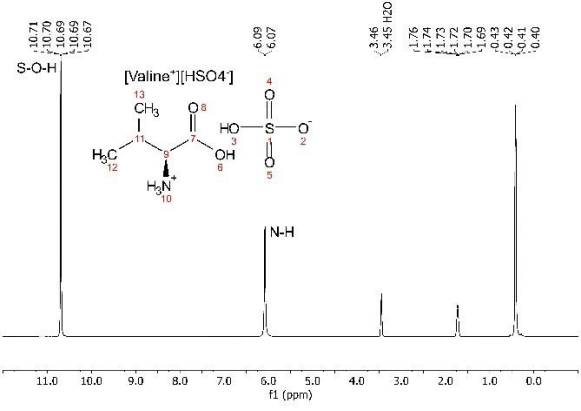
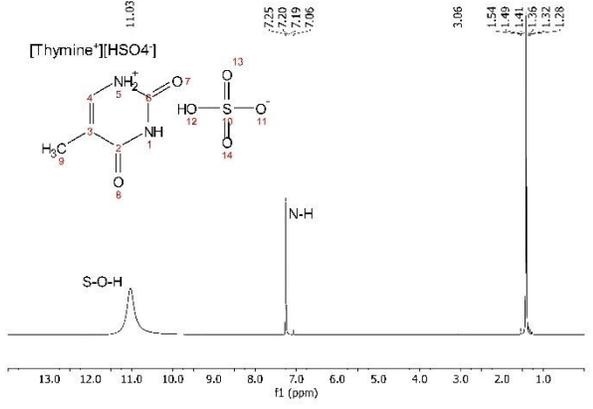
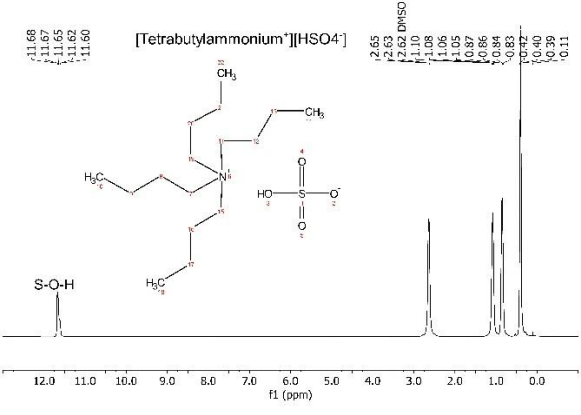
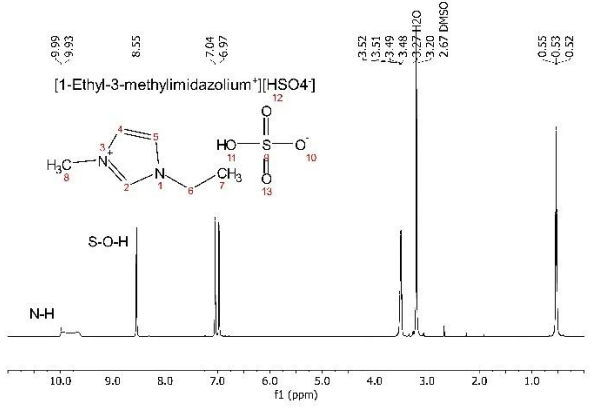
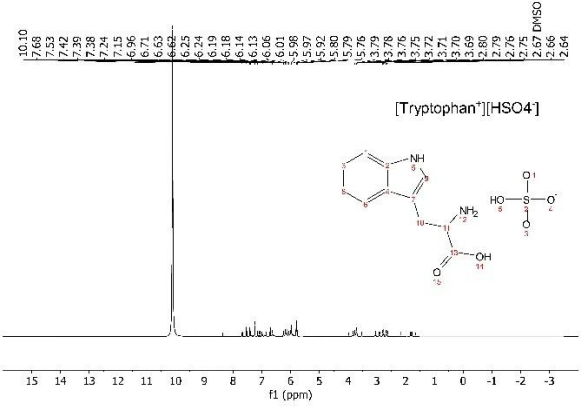

Figure S3. $^1$H NMR spectra of selected ionic liquids formed from various N-containing organics dissolved in concentrated sulfuric acid (98% $H_2SO_4$ w/w, the rest water) in equimolar (i.e., 1:1) concentration after the evaporation procedure. The spectra were recorded at 80 °C and 500.18 MHz with DMSO-$d_6$ as an external lock solvent. Note that even the unstable and reactive organics, like tryptophan, form ionic liquids with concentrated sulfuric acid.

Table S4. $^1$H NMR peaks of selected ionic liquids with signals assigned to N-H and S-O-H protons.

| Sample | δN-H(ppm) | δSO-H(ppm) |
|---|---|---|
| [glycine$^+$][HSO$_4^-$] | 6.18 | 10.26 |
| [glycine$^+$][HSO$_4^-$][H$_2$SO$_4$] | 6.03 | 10.35 |
| [glycine$^+$][HSO$_4^-$][H$_2$SO$_4$]$_2$ | 5.95 | 10.37 |
| [alanine$^+$][HSO$_4^-$] | 6.25 | 10.39 |
| [cysteine$^+$][HSO$_4^-$] | 6.16 | 10.46 |
| [valine$^+$][HSO$_4^-$] | 6.09 | 10.69 |
| [tryptophan products$^+$][HSO$_4^-$] | - | 10.10 |
| [thymine$^+$][HSO$_4^-$] | 7.20 | 11.03 |
| [tetrabutylammonium$^+$][HSO$_4^-$] | - | 11.65 |
| [1-ethyl-3-methylimidazolium$^+$][HSO$_4^-$] | ~10.00 | ~10.00 |
| [glycine$^+$][HSO$_4^-$][thymine$^+$][HSO$_4^-$] | 6.17, 7.15 | 10.22 |
| [ala-gly$^+$][HSO$_4^-$] | 6.30 | 10.51 |

Apart from amino acids, we also tested simple peptides like the alanylglycine (Ala-Gly) dipeptide for their ionic liquid formation capability (Figure S4). We note that Ala-Gly is stable in 98% sulfuric acid (the rest water) at room temperature for many months (7). Ala-Gly forms ionic liquid after heating to 80 °C for 24h followed by evaporation of excess sulfuric acid (see Section S2 for ionic liquid formation method and Section S6).

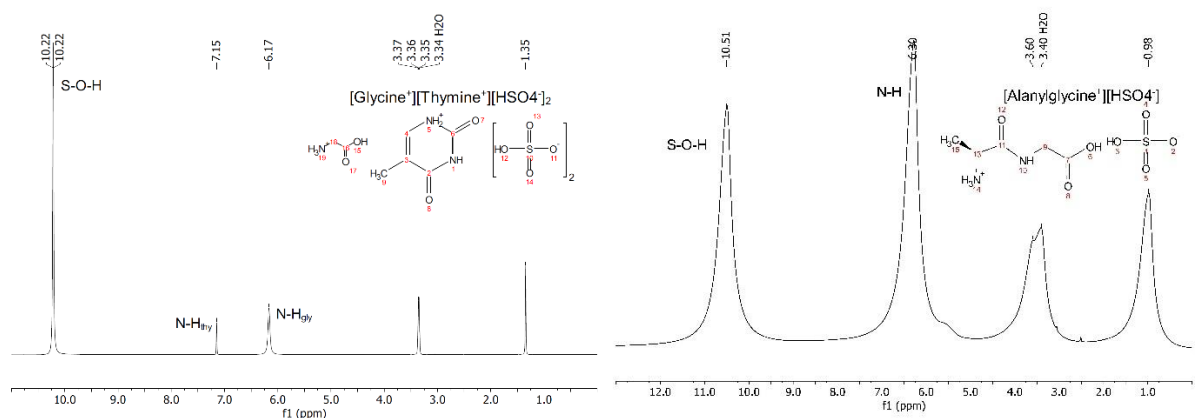

Figure S4. $^1$H NMR spectra of ionic liquid formed from a mixture of glycine and thymine in concentrated sulfuric acid (98% $H_2SO_4$ by weight, the rest water) in molar ratio of 1:1:2 of glycine:thymine:$H_2SO_4$ (left panel) and alanylglycine dipeptide ionic liquid (right panel). The spectra were recorded at 80 °C and 500.18 MHz with DMSO-$d_6$ as an external lock solvent.

**Fourier-Transform Infrared (FTIR) Spectroscopy**

We used FTIR spectroscopy to characterize selected ionic liquids (Figure S5) following the example of [glycine$^+$][HSO$_4^-$] ionic liquid described in the main text (Figure 3). We compared the IR spectra of [glycine$^+$][HSO$_4^-$] ionic liquid with the IR spectra of pure glycine and pure 98% w/w H$_2$SO$_4$ (the rest water) respectively. The carboxylate ion [COO$^-$] stretching in pure glycine appears between 1400 and 1600 cm$^{-1}$. In the ionic liquid, these bands shift to a higher wavenumber corresponding to [COOH] group vibration. The pure H$_2$SO$_4$ has a prominent peak at 1345 cm$^{-1}$, corresponding to S-OH bending vibrations. These vibrations are not prominent in the ionic liquid due to the strong ionic bonds. The [glycine$^+$][HSO$_4^-$] ionic liquid IR spectra are consistent with the IR spectra of hydrogen sulfate ionic liquids reported in the literature (8). Other tested ionic liquids exhibit similar features characteristic for hydrogen sulfate ionic liquids (Table S5).

This paragraph is repeated from the main manuscript and expanded upon. We obtained the infrared spectra of ionic liquids and pure forms of the constituents using a Bruker Alpha II FTIR spectrometer with a Diamond Crystal ATR (Attenuated Total internal Reflectance) accessory. We acquired 64 scans for each sample with 2 cm$^{-1}$ resolution. For each measurement of the ionic liquids, about ~5 µL liquid was transferred to the ATR crystal using a pipette. The liquid was exposed to air for 5-10 minutes while the spectra were obtained. Since the ionic liquids are highly hygroscopic (see Figure S11), this exposure to air in addition to variations in humidity during storage could lead to some absorption of water in the ionic liquids. This is a possible explanation for the shoulder around 3500 cm$^{-1}$ seen in a few of the IR spectra.

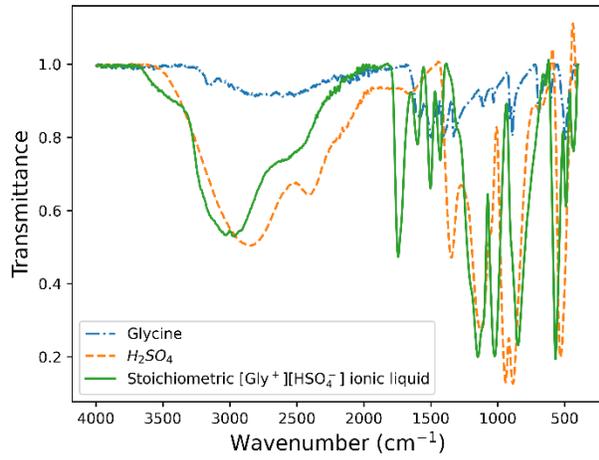
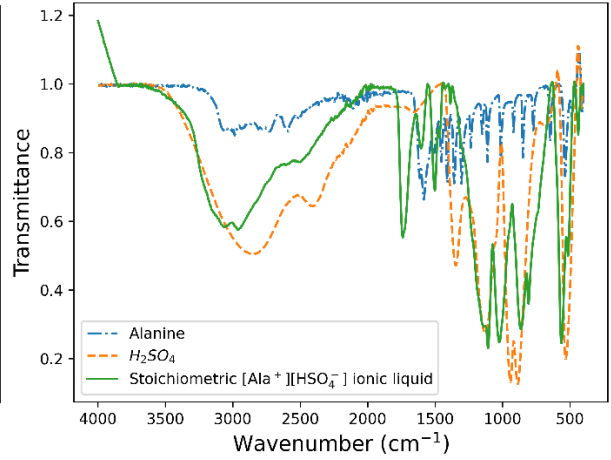
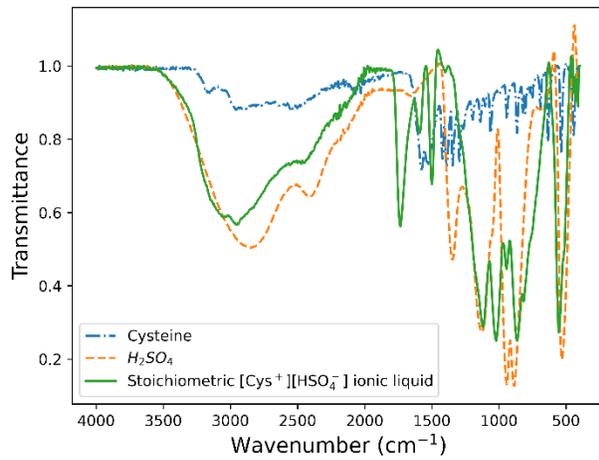
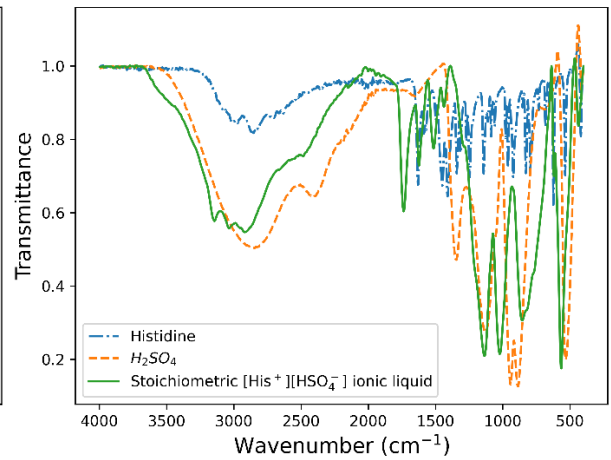
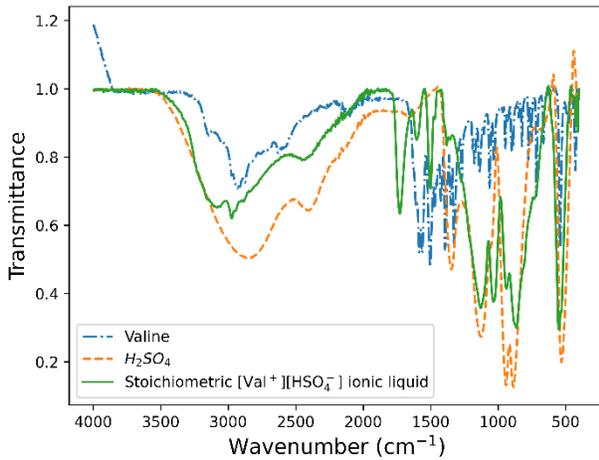
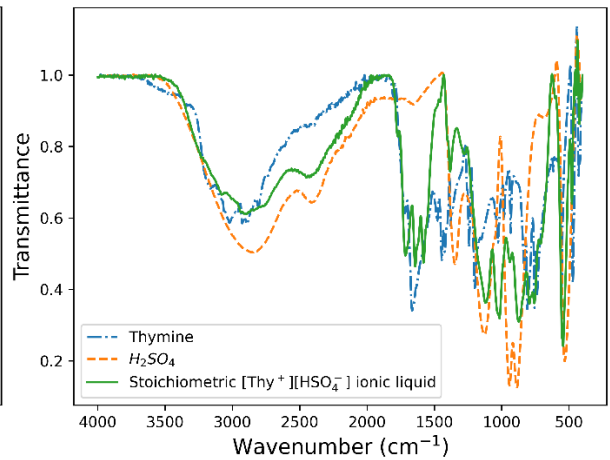

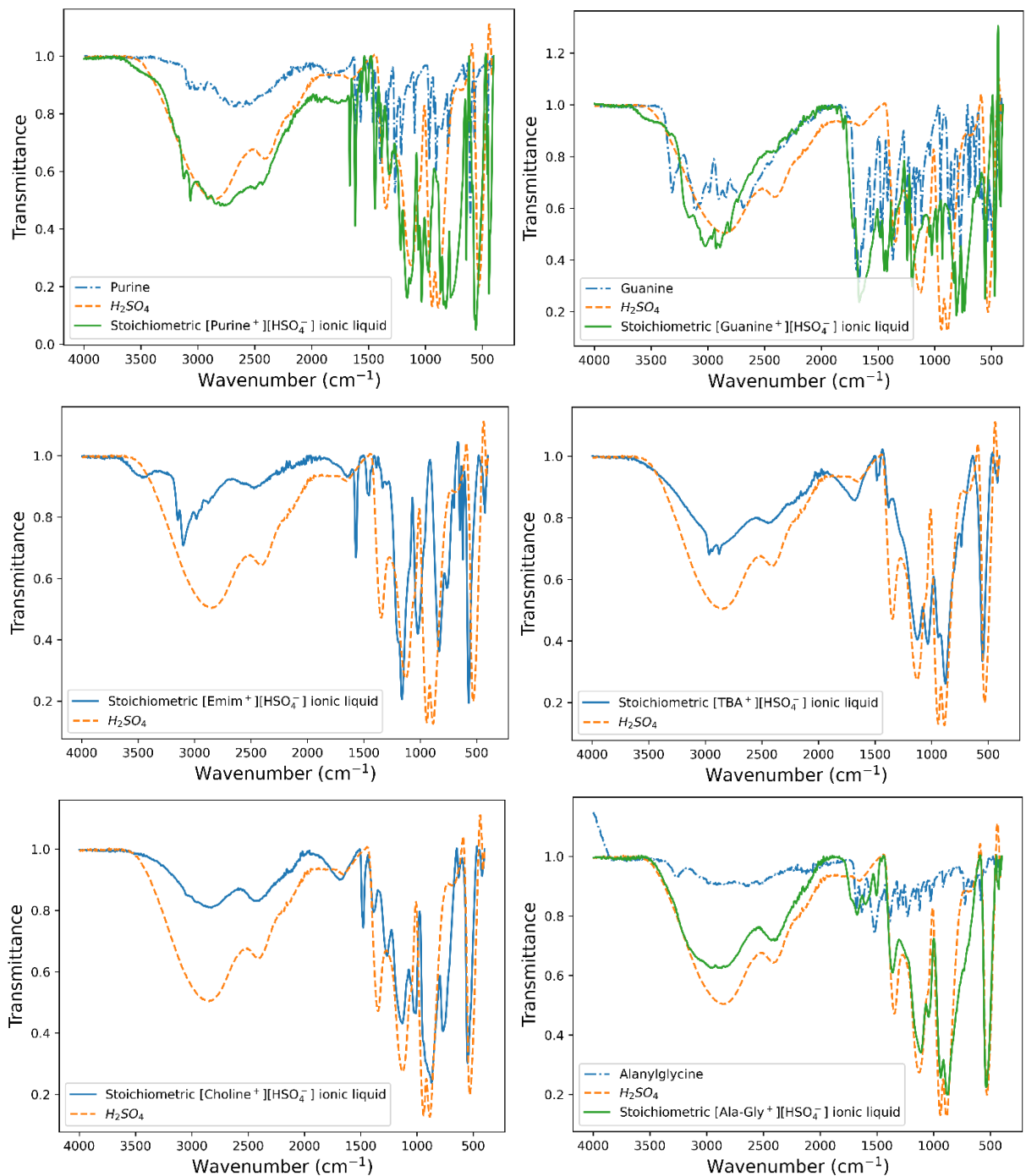

Figure S5. The FTIR spectra of selected ionic liquids formed from various N-containing organics dissolved in concentrated sulfuric acid (98% $H_2SO_4$ by weight, the rest water), in equimolar (i.e., 1:1) concentration, after the evaporation procedure. The y-axis is transmittance and the x-axis is wavenumber in cm$^{-1}$. (Emim is 1-ethyl-3-butylimidazolium; TBA is tetrabutylammonium).

Table S5. FTIR spectra peak assignments for the generated ionic liquids and $H_2SO_4$, compared with the peaks from [2].

| Sample | S-O | | S=O | | | S-OH |
|---|---|---|---|---|---|---|
| | [$HSO_4^-$] | $H_2SO_4$ | [$HSO_4^-$] | [$HSO_4^-$] | $H_2SO_4$ | $H_2SO_4$ |
| $H_2SO_4$ | 884.03 | 940.36 | 1047.10 | - | 1123.30 | 1345.04 |
| [glycine$^+$][$HSO_4^-$] | 852.01 | - | 1020.30 | 1139.57 | - | - |
| [glycine$^+$][$HSO_4^-$][$H_2SO_4$] | 853.46 | - | 1015.37 | - | 1103.47 | - |
| [glycine$^+$][$HSO_4^-$][$H_2SO_4$]$_2$ | 877.11 | 939.56 | 1027.32 | - | 1101.25 | - |
| [histidine$^+$][$HSO_4^-$] | 851.30 | - | 1022.78 | 1135.35 | - | - |
| [alanine$^+$][$HSO_4^-$] | 861.19 | - | 1025.21 | - | 1107.00 | - |
| [cysteine$^+$][$HSO_4^-$] | 865.19 | 940.97 | 1020.96 | 1120.09 | - | - |
| [valine$^+$][$HSO_4^-$] | 864.89 | - | 1034.35 | 1130.21 | - | - |
| [thymine$^+$][$HSO_4^-$] | 853.30 | - | 1011.00 | | 1106.00 | |
| [purine$^+$][$HSO_4^-$] | 864.95 | - | 1034.02 | 1160.79 | - | 1320.67 |
| [guanine$^+$][$HSO_4^-$] | 809.41 | 933.44 | 1026.01 | 1200 | - | - |
| [1-ethyl-3-methylimidazolium$^+$][$HSO_4^-$] | 831.74 | - | 1023.56 | 1163.25 | - | - |
| [tetrabutylammonium$^+$][$HSO_4^-$] | 875.29 | - | 1032.07 | 1127.38 | - | - |
| [choline$^+$][$HSO_4^-$] | 871.27 | - | 1011.78 | 1128.69 | - | 1383.10 |
| [alanylglycine$^+$][$HSO_4^-$] | 875.78 | 936.86 | 1045.31 | - | 1111.18 | 1363.67 |
| [glycine$^+$][$HSO_4^-$][thymine$^+$][$HSO_4^-$] | 857.59 | 941.01 | 1021.92 | | 1106.10 | - |
| Organic mixture 1 | 832.67 | - | 1022.32 | 1145.59 | - | - |
| $H_2SO_4$ (8) | 885.60 | 942.70 | 1047.10 | - | 1125.50 | 1347.50 |
| [Hmpyr$^+$][($HSO_4$)($H_2SO_4$)0.5] (8) | 844.20 | - | 1020.50 | 1150.80 | - | - |

The comparison of the IR spectra of the [glycine$^+$][$HSO_4^-$] ionic liquid with the spectra of the liquid that arises from evaporating glycine (280 mg) and acetic acid (300 µL) dissolved in 450 µL 98% w/w $H_2SO_4$ (the rest water) show that the two spectra match (Figure S6), confirming that acetic acid evaporates from the mixture and does not inhibit the formation of [glycine$^+$][$HSO_4^-$] ionic liquid. We additionally tested a mixture of glycine and thymine in concentrated sulfuric acid (in molar ratio 1:1:2 of gly:thy:$H_2SO_4$) and found that such a mixture also formed ionic liquid (Figure S7). In contrast to the pure [thymine$^+$][$HSO_4^-$] salt that is solid at room temperature (but liquid at 100 °C) the mixture of glycine and thymine forms a room temperature ionic liquid.

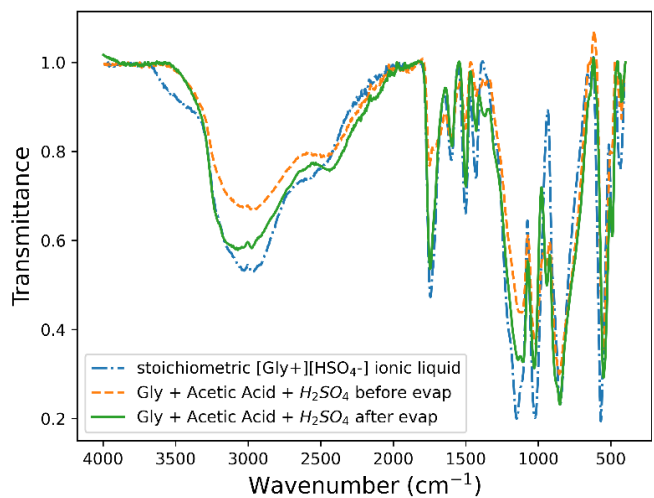

Figure S6. FTIR spectrum of [glycine+][HSO4-] ionic liquid (blue) compared with the ionic liquid generated from a glycine, acetic acid and $H_2SO_4$ mixture (orange). The y-axis is transmittance and the x-axis is wavenumber in cm$^{-1}$. The two spectra match, showing that acetic acid evaporates from the mixture and does not inhibit the formation of [glycine+][HSO4-] ionic liquid.

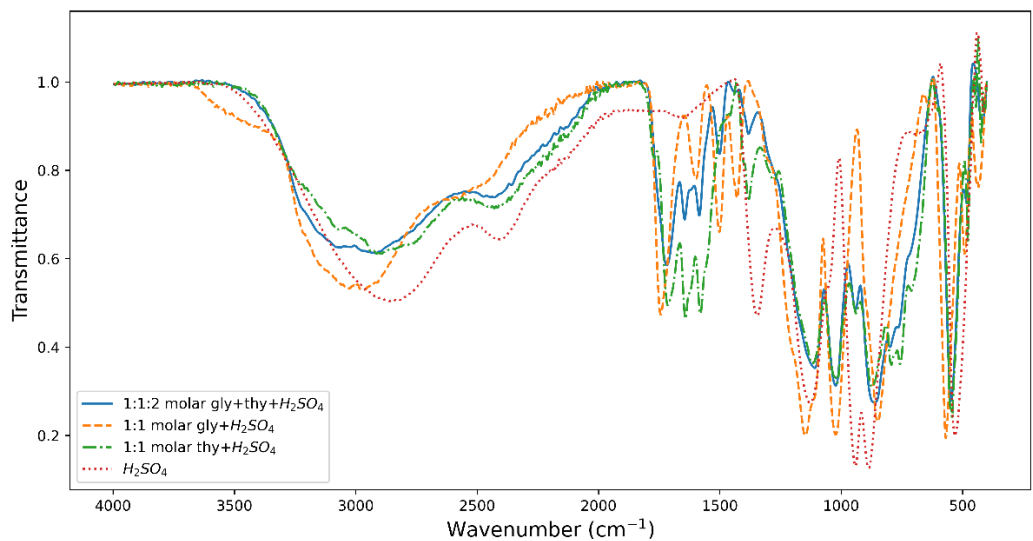

Figure S7. FTIR spectrum of a mixture of glycine and thymine-based ionic liquids with concentrated sulfuric acid. The y-axis is transmittance and the x-axis is wavenumber in cm$^{-1}$. The FTIR spectra of neat ionic liquids [glycine+][HSO4-] (orange spectra), [thymine+][HSO4-] (green spectra) compared to [glycine+][thymine+][HSO4-] (blue spectra), a mixture of the two organic components with concentrated sulfuric acid (98% $H_2SO_4$ w/w, the rest water) in molar ratio of 1:1:2 of glycine:thymine:$H_2SO_4$. In contrast to [thymine+][HSO4-] the mixture of glycine and thymine forms a room temperature ionic liquid.

Other tested mixtures include a mixture of glycine (114 mg) and sugar (112 mg) in 400 µL concentrated sulfuric acid (Figure S8). Another mixture contained N-containing organics, carboxylic acids, and hydrocarbons. The mixture was prepared by mixing glycine (10 mg), thymine (10 mg), stearic acid (50 mg), paraffin (wax) (5 mg), and naphthalene (5 mg), in 1 mL concentrated sulfuric acid (98% w/w $H_2SO_4$ the rest water) (Figure S9). The sugar, wax, and naphthalene were household items. The spectra before evaporation has dominant $H_2SO_4$ features and matches closely with the pure $H_2SO_4$ spectra. The spectra after evaporation shows a rightward shift in the S-O vibration region corresponding to the $HSO_4^-$ anion, consistent with other ionic liquids (Table S5).

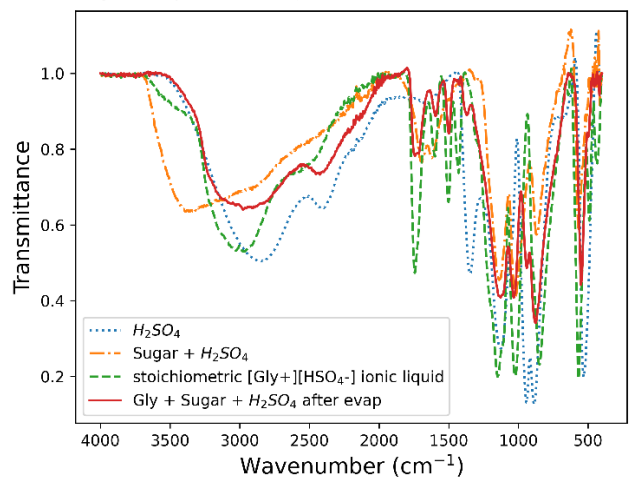

Figure S8. FTIR spectrum of [glycine$^+$][HSO$_4^-$] ionic liquid (green) compared with the ionic liquid generated from a glycine, sugar and $H_2SO_4$ mixture (red). The y-axis is transmittance and the x-axis is wavenumber in cm$^{-1}$. The bands corresponding to [glycine$^+$] (1400-1500 cm$^{-1}$ region) and [HSO$_4^-$] ions (850-1100 cm$^{-1}$ region) match, showing that sugar does not inhibit the formation of [glycine$^+$][HSO$_4^-$] ionic liquid but reacts with the excess sulfuric acid.

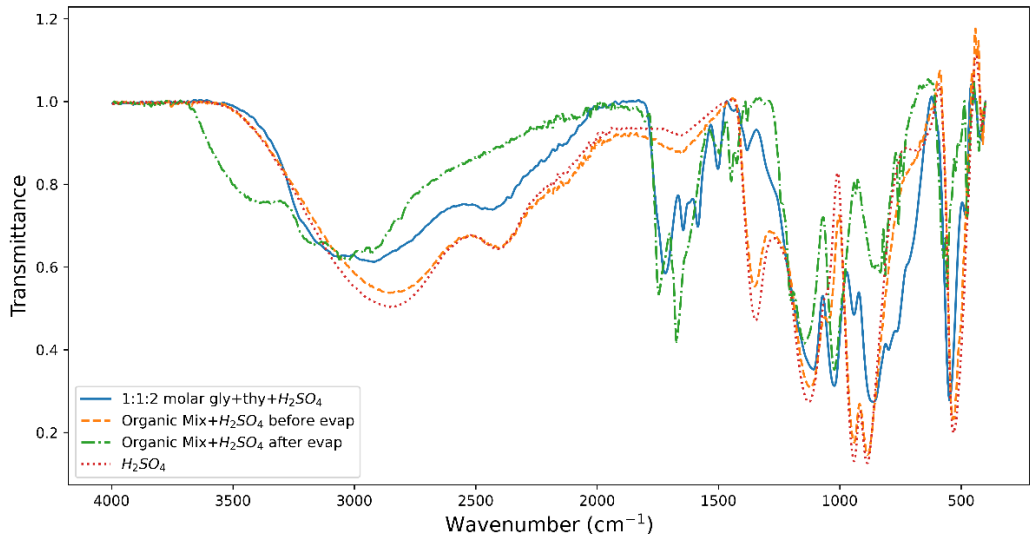

Figure S9. FTIR spectrum of an organic mixture (glycine, thymine, stearic acid, paraffin, naphthalene) in 98% w/w sulfuric acid, the rest water. The y-axis is transmittance and the x-axis is wavenumber in cm$^{-1}$. The IR spectra of the mixture before and after evaporation of the excess sulfuric acid differs significantly. The evaporation of the complex mixture of organics also results in the formation of ionic liquid.

## Thermogravimetric Analysis (TGA)

We used the TGA method to determine the onset temperature of thermal decomposition of selected ionic liquids. We found that the tested ionic liquids have various decomposition temperatures depending on component molecules and mixtures, but in all cases the decomposition temperatures are higher than about 180–190 ℃ (Figure S10 and Table 1 in the main text).

This paragraph is repeated from the main manuscript. About 1-2 mg of neat ionic liquid was loaded onto a platinum pan. The samples were gradually heated from 30 °C to 600 °C with a rate of 10 °C /min under the flow of nitrogen. The weight change was measured using TA Instruments Discovery TGA 5500.

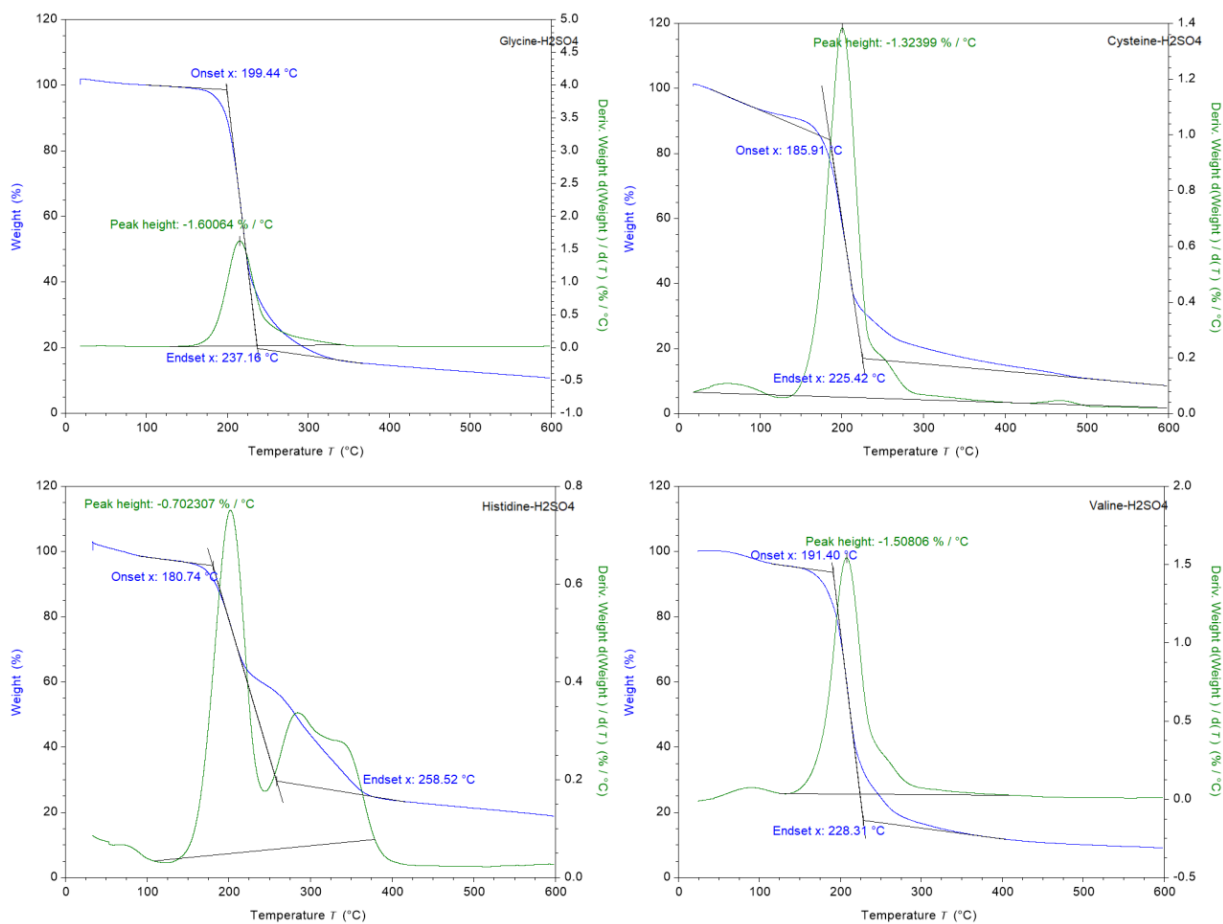

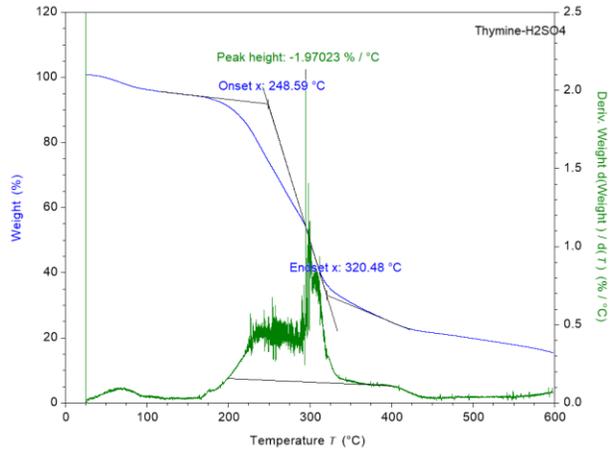
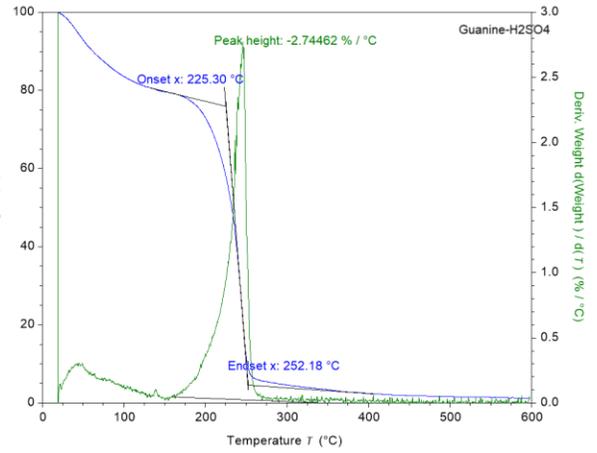
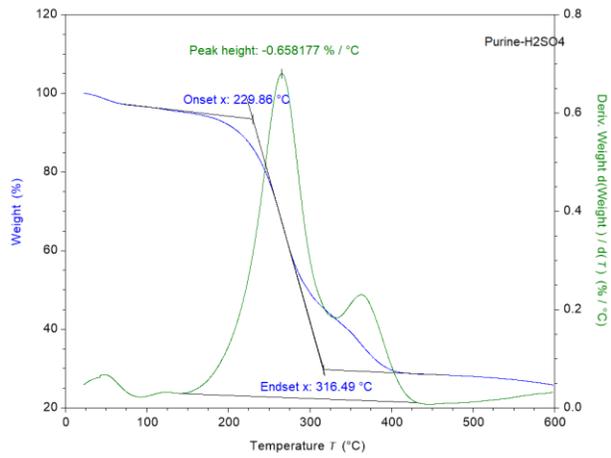
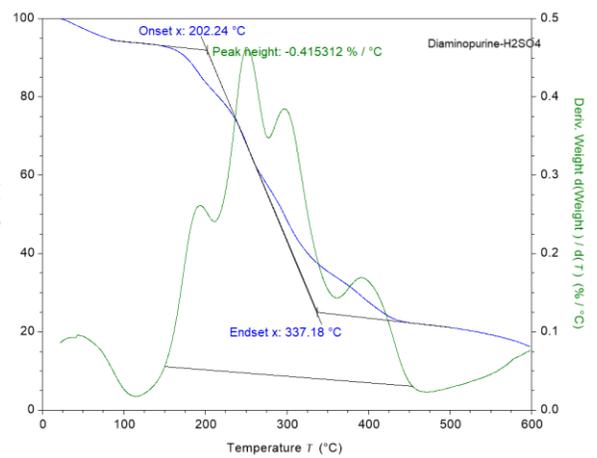
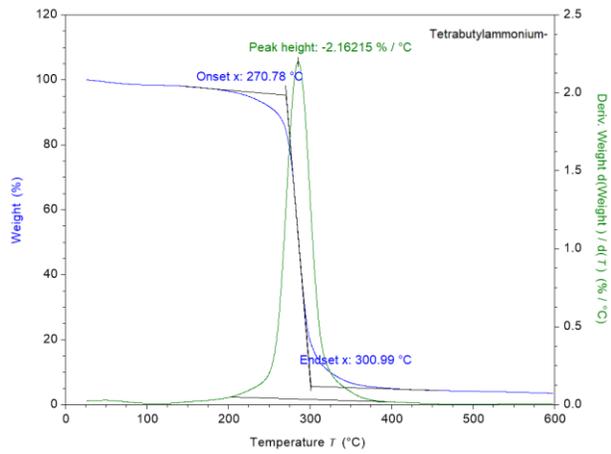
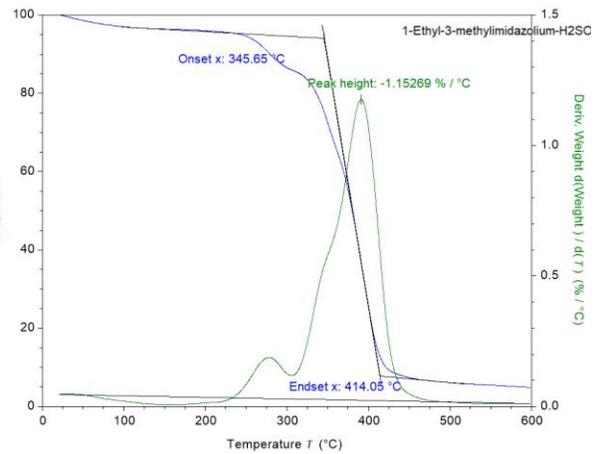

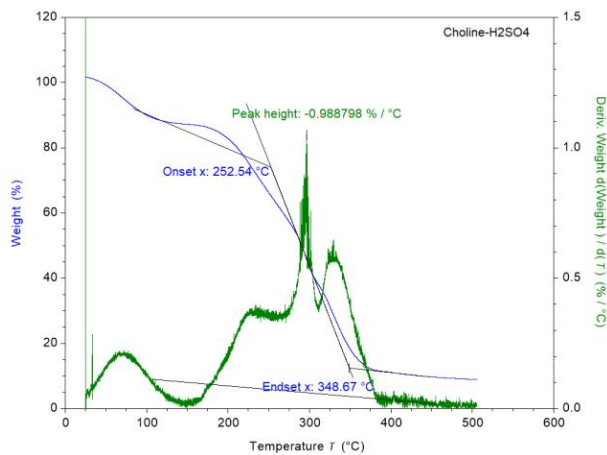

Figure S10. Thermogram curves of selected equimolar ionic liquids containing various N-containing organic molecules. The x-axis is temperature and the left-side y-axis is sample weight % remaining, while the right-side y-axis is first derivative of the weight % with temperature. The temperatures corresponding to the onset, peak, and endset of decomposition are plotted on the curves. The tested ionic liquids have various decomposition temperatures depending on the component organic molecule, but in all cases the decomposition temperature is higher than about 180–190 °C (see also Table 1 in the main text).

## $^{31}$P NMR Acceptor Number Determination

The hydrogen sulfate ionic liquids are acidic in nature (8). We use the Gutmann Beckett Acceptor Number (AN) $^{31}$P NMR method to determine the Lewis acidity of selected ionic liquids (9, 10). We show that, as expected, all tested hydrogen sulfate ionic liquids are quite acidic. The [tetrabutylammonium$^+$][HSO$_4^-$] ionic liquid is the least acidic of the tested substances (AN = 98) (Figure S11 and Table S6). In all cases, however, the acidity of ionic liquids is lower than the acidity of sulfuric acid (AN = 122).

This paragraph is repeated from the main manuscript and expanded upon. We used triethylphosphine oxide (TEPO) as the probe for $^{31}$P NMR. Approximately 50 mg of TEPO was added to ~1 ml of the ionic liquid and incubated at 80 °C to allow mixing. The samples were transferred into 5 mm NMR tubes, each with a capillary tube insert containing DMSO-d$_6$ as external lock. $^{31}$P NMR spectra was obtained at 80 °C using a Bruker Avance Neo spectrometer operating at 500.18 MHz. We measured the spectra for H$_2$O and H$_2$SO$_4$ to establish the range of acidity within which the ionic liquid acidity must lie (Figure S11). Table S6 lists the $^{31}$P NMR shift of TEPO ($\delta$ET$_3$PO) and the corresponding acceptor number. The acceptor number is calculated by using the formula: AN = 2.218 × ($\delta$ET$_3$PO - 41)

We also performed additional AN estimates to confirm the complete evaporation of excess sulfuric acid and the formation of equimolar ionic liquid. We prepared an equimolar ionic liquid by directly adding equal molar amounts of glycine and H$_2$SO$_4$ (1400 mg of glycine and 1.037 ml of 98% w/w H$_2$SO$_4$ (the rest water), without extra H$_2$SO$_4$ and without the evaporation step. This equimolar mixture has an AN of 112.8, a value very similar to the AN value of [glycine$^+$][HSO$_4^-$] ionic liquid obtained by the evaporation of excess sulfuric acid (AN = 113.1). The obtained AN values further confirm efficient removal of the excess sulfuric acid and the formation of the equimolar

[glycine$^+$][HSO$_4^-$] ionic liquid. The AN values also confirm the high acidity of the [glycine$^+$][HSO$_4^-$] ionic liquid.

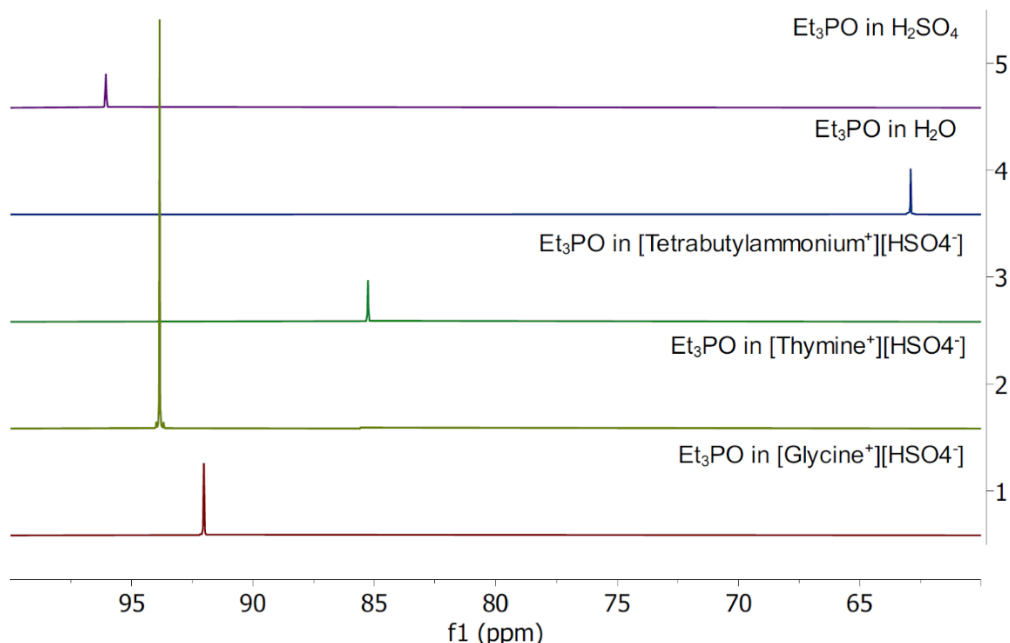

Figure S11. $^{31}$P NMR spectra of triethylphosphine oxide (TEPO) dissolved in selected ionic liquids at 80 °C and 500.18 MHz, with DMSO-d$_6$ external lock. The [tetrabutylammonium$^+$][HSO$_4^-$] ionic liquid is the least acidic of the tested ionic liquids.

Table S6. The $^{31}$P NMR shift of TEPO and corresponding acceptor number of select ionic liquids, 98% w/w concentrated H$_2$SO$_4$ (the rest water), and pure H$_2$O.

| Sample | δET$_3$PO (ppm) | Acceptor Number |
|---|---|---|
| [glycine$^+$][HSO$_4^-$] | 92 | 113.1 |
| [thymine$^+$][HSO$_4^-$] | 93.8 | 117.1 |
| [tetrabutylammonium$^+$][HSO$_4^-$] | 85.2 | 98.0 |
| H$_2$O | 62.9 | 48.6 |
| 98% w/w H$_2$SO$_4$ in H$_2$O | 96 | 122.0 |

## Hygroscopicity

Ionic liquids are very hygroscopic. The [glycine$^+$][HSO$_4^-$] ionic liquid accumulates a significant amount of water from the surrounding air when left in an open container. A ~1.4 gram of ionic liquid gained 0.34 gram in mass in 14 days (Figure S12). For example, to test how much water the [glycine$^+$][HSO$_4^-$] ionic liquid accumulates, we measured the initial weight of the sample right after removing it from the glass vial and placed the sample in an open vial (open to ambient lab conditions). The weight of the sample was measured in regular intervals over 14 days. Due to hygroscopicity of ionic liquids, we store them in airtight glass vials.

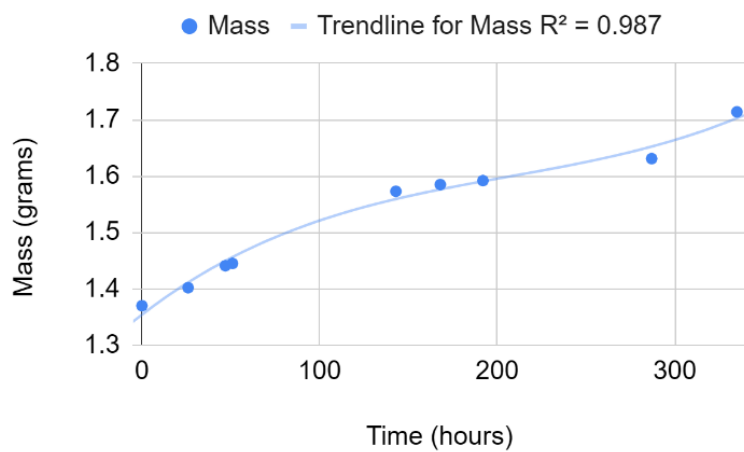

Figure S12. The increase in the mass of the [glycine+][HSO$_4$-] ionic liquid due to accumulation of water from the surrounding air. The y-axis is the total mass of the ionic liquid and the x-axis is the time the vial containing an ionic liquid has been left open (in hours). The mass increases monotonically. The non-linear behavior can be attributed to the variation in daily humidity which could cause differences in the rate of accumulation of water over time.

## S4. Discussion on Ionic Liquid Formation

### Formation of Ionic Liquid with Varying Concentration of Sulfuric Acid in Water

We show that the starting concentration of sulfuric acid does not affect the formation of ionic liquids. We mixed a fixed quantity of glycine in varying concentrations of sulfuric acid in water by weight (0%, 5%, 25%, 50%, 75%, 95%) (Table S7). We evaporated the samples in our custom-built vacuum chamber at 100 °C (+/- 5 °C) and 0.01 mbar pressure for 12 hours, as described in Section S2. We measured the FTIR spectra of the sample before and after evaporation (Figure S11). In the diluted samples, the O-H peak corresponding to $H_2O$ is clearly visible. The post-evaporation spectra of all the samples with non-zero concentration of $H_2SO_4$ have no H2O OH peak and are consistent with the equimolar ionic liquid [glycine$^+$][HSO$_4^-$]. After evaporation, the sample with 0% $H_2SO_4$ leaves a white dry powder of glycine.

Table S7. Quantities used in the experiment to prepare mixtures of glycine in varying $H_2SO_4$ concentration by weight in $H_2O$.

| Sample No. | $H_2SO_4$ Conc. in $H_2O$ w/w% | Glycine (mg) | $H_2SO_4$ (µL) | $H_2O$ (µL) |
|---|---|---|---|---|
| 1 | 0 | 103 | 0 | 1000 |
| 2 | 5 | 98 | 80 | 2830 |
| 3 | 25 | 100 | 80 | 450 |
| 4 | 50 | 105 | 80 | 150 |
| 5 | 75 | 99 | 80 | 50 |
| 6 | 95 | 96 | 80 | 0 |

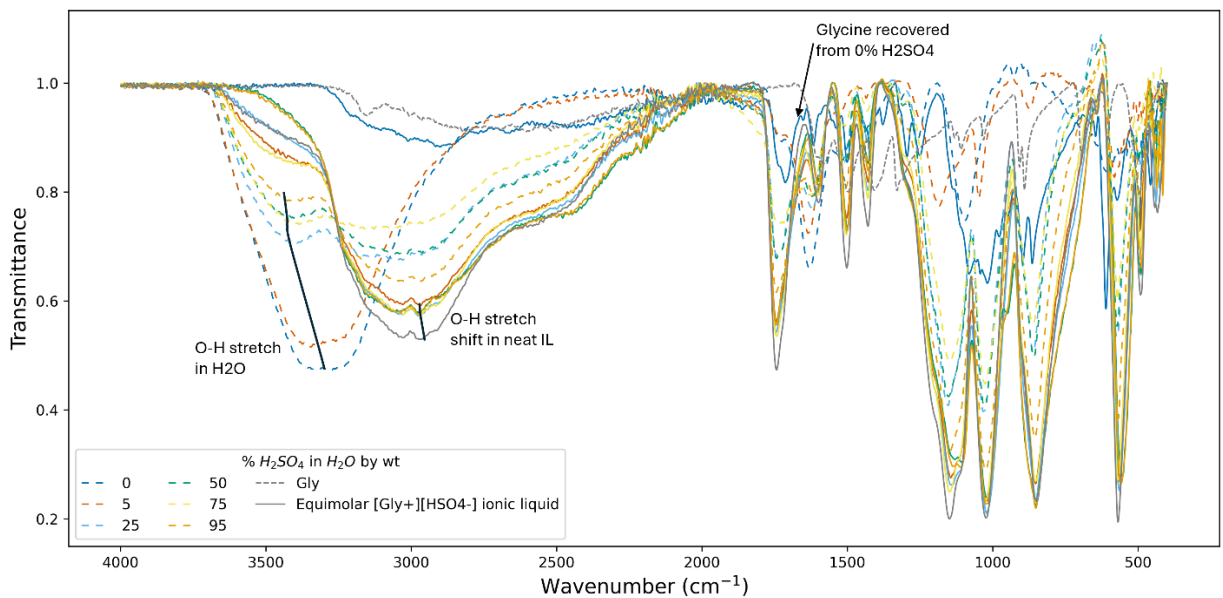

Figure S13. FTIR spectra of ionic liquids formed from mixtures of glycine dissolved in various concentrations of H$_2$SO$_4$ in water. The y-axis is transmittance and the x-axis is wavenumber in cm$^{-1}$. The dashed lines are mixtures before evaporation and the same color solid lines are after evaporation. The FTIR spectra of the evaporated samples with non-zero concentration of H$_2$SO$_4$ are consistent with the equimolar ionic liquid [glycine$^+$][HSO$_4^-$], showing the formation of ionic liquids no matter the starting concentration of sulfuric acid.

## Formation of Ionic Liquid with Varying Concentration of Glycine in Sulfuric Acid

Ionic liquids can exist at various molar ratios of cationic and anionic components. Compared to equimolar (1:1) ionic liquids, non-equimolar ionic liquids form by combination of neat ionic liquids with a molar excess of the cationic or anionic species (11). In our exoplanet scenario the abundance of anionic species, i.e. sulfuric acid, will dominate over the available amounts of cationic, organic molecules. Once the excess concentrated sulfuric acid evaporates, the ionic liquid is produced in the equimolar ratio of the cationic and ionic components. Nonetheless, we investigate non-equimolar ratios of the [glycine$^+$][HSO$_4^-$] (1:2 and 1:3 molar ratios) to further characterize the properties of ionic liquids. We mixed glycine in three molar ratios (1:1, 1:2, 1:3) in concentrated sulfuric acid without evaporation (Section S2 and Table S6).

The spectral characterization of equimolar and non-equimolar ionic liquids has two points of relevance. First, the non-equimolar (1:2 and 1:3) ionic liquids still contain the anionic and cationic components but due to the excess moles of HSO$_4^-$ ions or H$_2$SO$_4$ species their FTIR and $^1$H NMR spectra are different than the equimolar sample. This difference implies that their physicochemical properties are different as well (Figures S14, S15, S16).

Second, the evaporated diluted solution of glycine (20 mM, 0.001:1 molar ratio) dissolved in 98% w/w sulfuric acid (the rest water) reaches the equimolar ratio, thus forming the [glycine$^+$][HSO$_4^-$] ionic liquid (Figure S16). We show that when a non-equimolar ionic liquid is subjected to temperature and pressure conditions that allow evaporation of sulfuric acid, the excess moles of

$H_2SO_4$ evaporate from the solution resulting in an equimolar ionic liquid containing just $HSO_4^-$ anions and the positively charged N-containing organic.

Additionally, we tested the ionic liquid formation with the smallest volume that we could obtain FTIR spectra for, on order of 200 nano liters (nL). We used 300 µg of glycine mixed with 20 µL of concentrated sulfuric acid (98% by weight in water) and evaporated the excess sulfuric acid using our custom evaporator. The FTIR spectra post-evaporation matches the [glycine$^+$][$HSO_4^-$] ionic liquid (Figure S16).

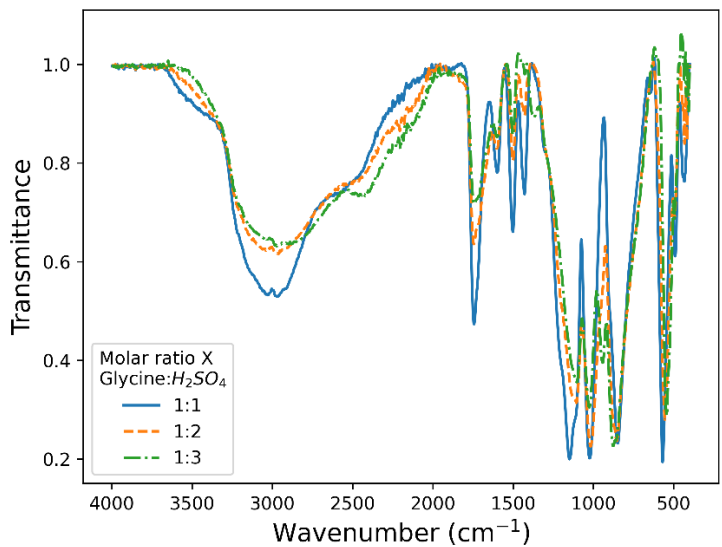

Figure S14. FTIR spectra of glycine in concentrated sulfuric acid (98% $H_2SO_4$ by weight, the rest water) at three molar ratios, 1:1, 1:2, and 1:3. The y-axis is transmittance and the x-axis: wavenumber in cm$^{-1}$. The IR spectra of the equimolar [glycine$^+$][$HSO_4^-$] ionic liquid mixture (blue) is distinct from the 1:2 (orange) and 1:3 (green) spectra allowing for differentiation between equimolar and non-equimolar 1:2 and 1:3, ionic liquids. The FTIR of the 1:2 and 1:3 ionic liquids have a wide O-H stretch in the 3200-2300 cm$^{-1}$ region, whereas the 1:1 variant has a narrower O-H region 2500-3200 cm$^{-1}$ and is shifted as compared to its 1:2 and 1:3 counterparts.

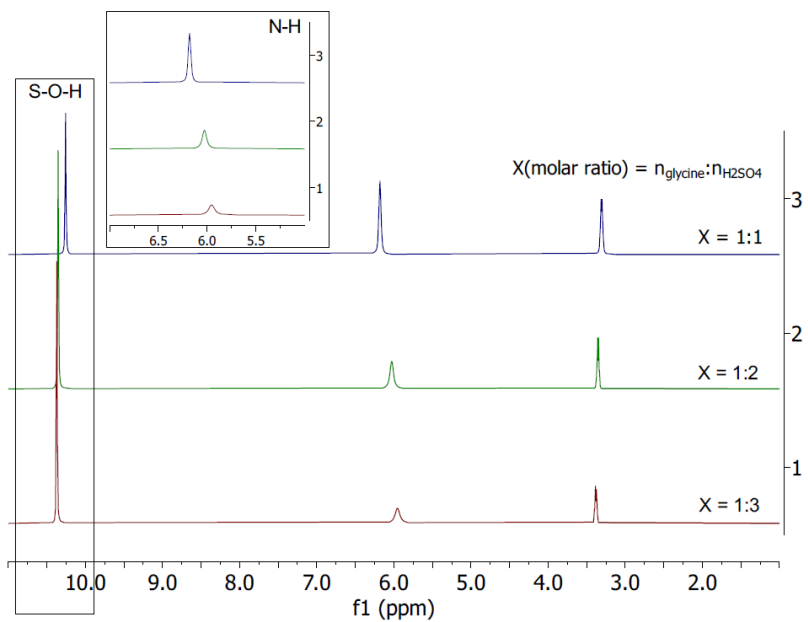

**Figure S15**. $^1$H NMR spectra of glycine in concentrated sulfuric acid (98% $H_2SO_4$ w/w, the rest water) at three molar ratios. The spectra are recorded at 80 °C and 500.18 MHz with DMSO-$d_6$ as an external lock solvent. The molar composition of [glycine$^+$][HSO$_4^-$] ionic liquid variants are: 1:1 (blue spectra), 1:2 (green spectra), and 1:3 (red spectra). The $^1$H NMR spectra of [glycine$^+$][HSO$_4^-$] ionic liquid produced by evaporation of the excess acid is consistent with an equimolar ratio ionic liquid and differs from the spectra of non-equimolar ionic liquids (1:2 and 1:3). The SO-H proton peak at 10.26 ppm is slightly shifted upfield in the 1:1 variant as compared to non-equimolar ionic liquids (1:2 and 1:3). Such an upfield shift agrees with previously reported spectra of hydrogen sulfate ionic liquids (8). Likewise, the peak corresponding to the amine protons of the glycine group, at 6.18 ppm, is slightly shifted downfield in the equimolar [glycine$^+$][HSO$_4^-$] ionic liquid, as compared to its 1:2 and 1:3 counterparts.

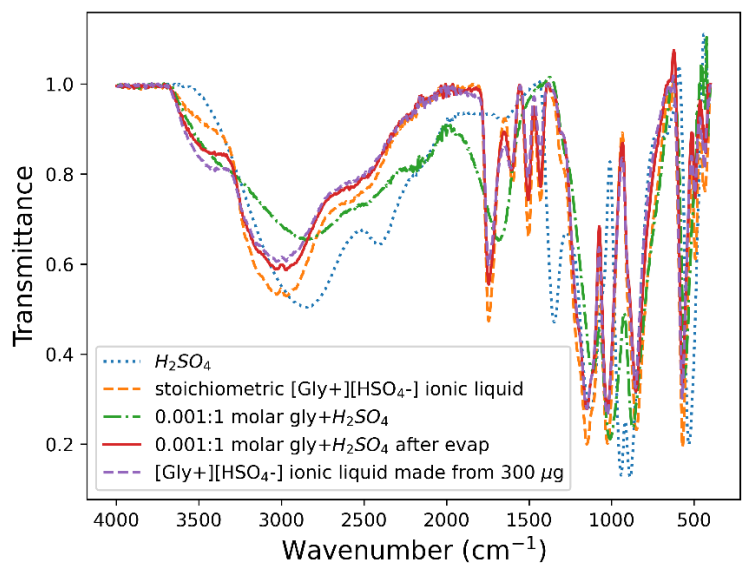

Figure S16. FTIR spectra of 20 mM glycine dissolved in 98% w/w $H_2SO_4$ (the rest water) before and after evaporation of excess sulfuric acid, compared to the equimolar [glycine$^+$][HSO$_4^-$] ionic liquid. The y-axis is transmittance and the x-axis is wavenumber in cm$^{-1}$. The scale of preparation, i.e. the amount of dissolved glycine, does not affect the formation and composition of the ionic liquid. The spectra of equimolar [glycine$^+$][HSO$_4^-$] ionic liquid look virtually identical to the spectra of ionic liquid formed from evaporated diluted solution of glycine (orange spectra, 20 mM, 0.001:1 molar ratio).

## Formation of Ionic Liquids on a Basalt Rock Surface

We show that ionic liquids form outside of pure laboratory conditions, on the surface of basaltic rocks, in a variety of conditions—as long as the excess $H_2SO_4$ is removed. In our experiments, the acid is removed by spreading through rock pores, reacting with the rock, or evaporation. As a relevant side note, concentrated sulfuric acid itself does not readily react with basalt rock surfaces under dessicated conditions (Figure S17). Also, glycine powder deposited on the surface of basaltic rocks readily dissolves in hot liquid concentrated sulfuric acid.

In all of the experiments we used grayish black trap rock from Ward's Science company. We placed two basaltic rocks in our vacuum chamber, with the glycine-$H_2SO_4$ mixture on one rock and only liquid $H_2SO_4$ on the second rock (as a control). The hot plate was set to 120 °C, which resulted in 80 °C (+/- 5 °C) temperature of the rocks (measured with an infrared thermometer). After 24 hours, we observed that the rock with glycine was coated with liquid, whereas the control rock was almost dry, showing that excess sulfuric acid evaporated or was absorbed by the rock. The ionic liquid that formed on the basaltic rock was stable to evaporation and further reactivity with the rock surface. We note that the vacuum chamber conditions were the same those used for ionic liquid production in the glass vials (see SI Section S2 and results in Figure S18, S19 and Figure 2 in the main text);

Additionally, we show that in the presence of N-containing organics, sulfuric acid forms ionic liquid before reacting with the basaltic rock—even under open, room-temperature, room-pressure conditions including in the presence of water. We tested the formation of ionic liquid on the basaltic rock without our custom evaporator under three conditions. In the first two conditions, glycine (28 mg) was transferred to drilled cavities in the rocks and 40 μL of 98% w/w $H_2SO_4$ (the rest water) was added. The first rock was placed in the open lab conditions (Figure S18a) and the second rock was placed inside a desiccated container (Figure S18b). In the third condition, glycine (28 mg) was mixed with 20 μL of water and 20 μL of 98% w/w $H_2SO_4$ (the rest water) in the rock cavity and placed inside the desiccated container (Figure S18 c). In all three cases, there was visible liquid inside the cavities after 40 hours and the FTIR spectra of the samples matched that of the [glycine$^+$][HSO$_4^-$] ionic liquid prepared in glass vials (Figure S19).

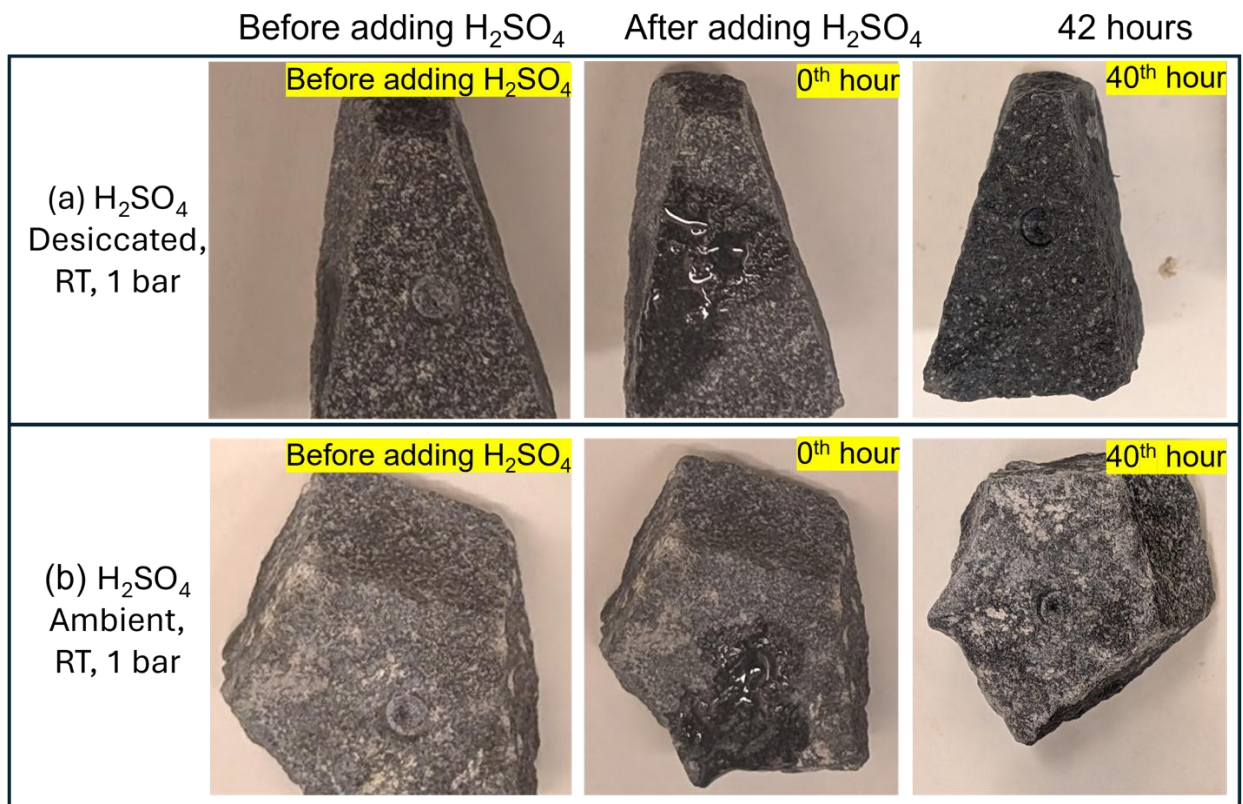

Figure S17. Concentrated sulfuric acid does not react with basalt in dry air. Top row: Basalt with 98% w/w $H_2SO_4$ (the rest water) stored in a desiccated jar to prevent interaction with water in the air. There is no apparent visible reaction of concentrated sulfuric acid (CSA) with the basalt rock in a desiccated environment. The rock remains coated in liquid. Bottom row: Basalt with 98% w/w $H_2SO_4$ (the rest water), stored in open lab conditions at all times. Water vapor from the air dilutes the acid and leads to reactivity of the diluted acid with the rock surface. Such weathering of rocks by aqueous, diluted, sulfuric acid solutions is well known (e.g., (12, 13)).

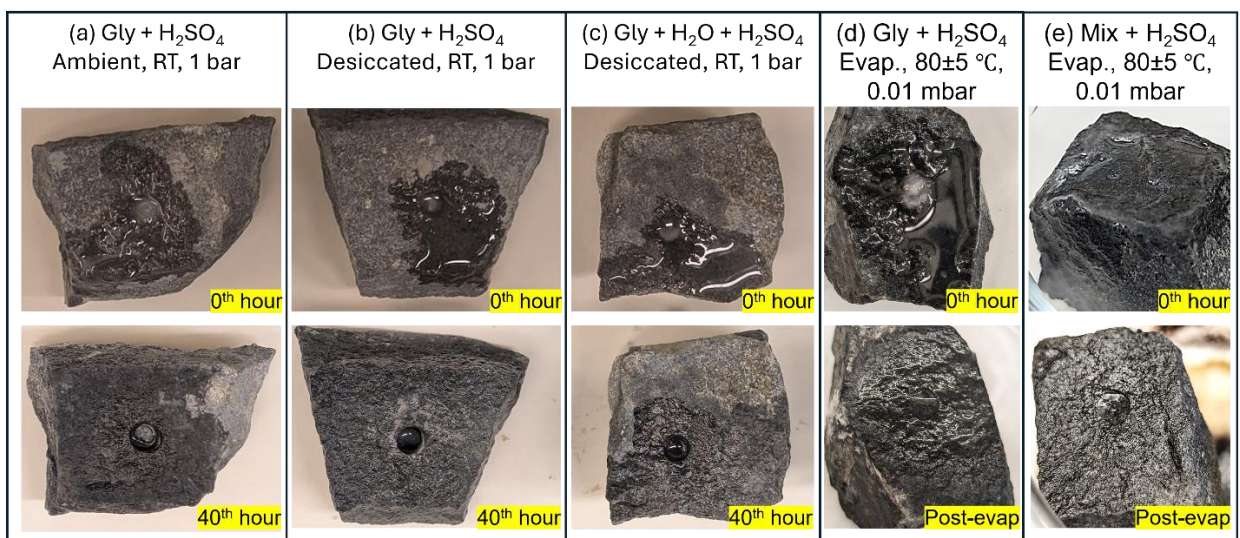

Figure S18. N-containing organics and $H_2SO_4$ on basalt form ionic liquids. Mix in (e) refers to a mixture of N-containing organic compounds—glycine, histidine, thymine, choline, tetrabutylammonium, and palmitic acid. Top row: 98% w/w $H_2SO_4$ (the rest water) and organic compounds right after mixing, with excess $H_2SO_4$. Bottom row: After 40 hours (a, b, c) or post evaporation (d, e). The bottom-row images show that ionic liquids form on rock surfaces after the excess $H_2SO_4$ is removed—either by spreading through rock pores, reacting with the rock, or evaporation.

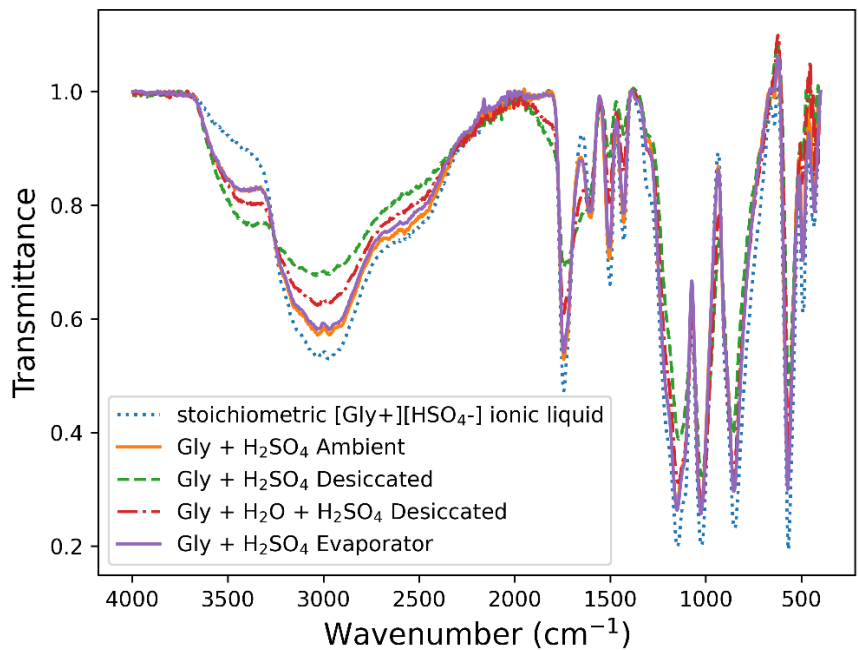

Figure S19. FTIR spectrum of [glycine+][HSO4-] ionic liquid (blue) compared with the ionic liquid generated from a glycine-$H_2SO_4$ mixture on the basalt rock under various conditions. The y-axis is transmittance and the x-axis is wavenumber in cm$^{-1}$. The spectra match, indicating that the liquid that forms on the surface of basaltic rock is [glycine+][HSO4-] ionic liquid. All the ionic liquid spectra look virtually identical, confirming the formation of [glycine+][HSO4-] ionic liquid on the surface of basaltic rock from different starting conditions. Note the y-axis offsets are due to differing volumes of liquid on the FTIR measuring crystal. The shoulder at ~3500 cm$^{-1}$ corresponds to the $v_{O-H}$ in $H_2O$ and is likely due to moisture absorbed from air, due to the highly hygroscopic nature of ionic liquids.

## Formation Pathways of Concentrated Sulfuric Acid

We capture an overview of concentrated sulfuric acid formation pathways on Earth, Venus, and Europa in Table S8.

Table S8. An overview of known planetary $H_2SO_4$ formation pathways.

| $H_2SO_4$ Formation Process | Location | Comments |
|---|---|---|
| UV-driven oxidation of $SO_2$, gas phase oxidation by OH (Earth) (14–17) or O (Venus) radicals (18–20). | Atmosphere: Venus, Earth | Formation of sulfuric acid clouds on Venus and Earth stratospheric aerosols. |
| Atmospheric oxidation of reduced organic sulfur compounds (21). | Atmosphere: Earth | Direct production ($SO_2$-independent) of sulfuric acid by atmospheric oxidation of reduced sulfur-containing organic compounds. |
| Oxidation of $SO_2$ in water cloud droplets (22, 23). | Atmosphere: Earth | Aqueous phase oxidation in cloud or fog droplets by dissolved $O_3$, $H_2O_2$, $NO_2$, and transition metal ions. |
| Volcanically produced in an oxidizing environment (24). | Surface: Earth | Sulfuric acid can be directly released by volcanoes in an oxidizing environment. |

| | | | | | | |
|---|---|---|---|---|---|---|
| Radiolysis and oxidation of surface sulfur deposits (25) (reviewed in (26)). | Surface: Europa | Sulfuric acid on the surface of Jupiter's moon Europa forms via radiolytic oxidation of sulfur deposits on Europa's icy shell. | | | | |
| $Fe_2O_3$-catalyzed oxidation of volcanically released $SO_2$ (27). | Surface: Earth | On Earth basaltic or mafic surface rocks containing $Fe_2O_3$ can catalyze $SO_2$ oxidation to $H_2SO_4$ in volcanically active environments rich in $SO_2$. | | | | |
| Heterogeneous oxidation of atmospheric $SO_2$ on the surfaces of various mineral dust particles: gibbsite (28), $MnO_x$: (29, 30), $Fe_2O_3$: (31, 32). | Atmosphere: Earth | Mineral dust in Earth's atmosphere can catalyze the formation of sulfuric acid. | | | | |

**Sulfuric Acid Evaporation Time Estimates**

We provide experimentally measured timescales (Table S9) for concentrated sulfuric acid evaporation for temperatures and pressures within the liquid portion of the sulfuric acid phase diagram (Figure 5 in the main manuscript). The point is that sulfuric acid can evaporate in reasonable times for a range of planetary temperature and pressure surface conditions, in order to form ionic liquids after dissolving organics.

Table S9. Time in months to evaporate 1 mL/cm$^2$ of pure concentrated sulfuric acid (98% w/w concentrated sulfuric acid, the rest water) at a range of temperatures and pressures. Values are scaled from evaporation measurements shown in the four right-most columns.

| Temperature (K) | Pressure (atm) | Time to Evaporate 1 mL/cm$^2$ (months) | Volume (μL) | Area (mm$^2$) | Time (min) | Quantity Spread Over Area (mL/cm$^2$) |
|---|---|---|---|---|---|---|
| 323 | 1 x 10$^{-5}$ | 1.3 | 1 | 52 | 108 | 0.001913 |
| 363 | 1 x 10$^{-3}$ | 0.4 | 1 | 59 | 30 | 0.001682 |
| 353 | 4 x 10$^{-5}$ | 0.1 | 1 | 19 | 31 | 0.005300 |
| 343 | 6 x 10$^{-5}$ | 1.0 | 2 | 55 | 163 | 0.003661 |
| 353 | 7 x 10$^{-4}$ | 0.7 | 2 | 36 | 160 | 0.005546 |
| 323 | 7 x 10$^{-4}$ | 1.9 | 2 | 65 | 255 | 0.003070 |
| 373 | 1.6 x 10$^{-3}$ | 0.2 | 2 | 34 | 45 | 0.005961 |
| 373 | 1 x 10$^{-5}$ | 0.3 | 2 | 24 | 90 | 0.008285 |

## Organics in the Solar System

We provide an overview of the known distribution of the organic material in the solar system in Table S10.

Table S10. Organic material in the solar system.

| Planetary Body | Organics Detected | Hypothesized Mechanism of Formation | Relevance to the Planet with Surface Ionic Liquids |
|---|---|---|---|
| Mercury | Potentially large abundance of complex organics of unknown composition (33). | Unknown, possibly originally delivered to Mercury's surface by comets. | Provides a precedent for localized abundance of complex organics, possibly as thick as 10 cm, on the surface of atmosphereless bodies. The composition of the organic deposits is unknown but it is likely a complex macromolecular material formed as a result of processing of simple organic molecules into dark complex organics by high-energy photons and particles. |
| Luna (the Moon) | Low abundance of diverse organic molecules in lunar regolith (34–36). | Combination of meteoritic or cometary infall to the lunar surface and terrestrial contamination from the Apollo program. | N/A |
| Mars | Aromatic and aliphatic Cl-containing organics, S-containing organics including thiophene, decane, undecane and dodecane (the last three are pyrolytic products of carboxylic acids) (37–43). | Unknown, possibly involving past aqueous environment. | N/A |
| Asteroids and meteoritic material | A variety of organic molecules including amino acids, nucleic acid bases, carboxylic acids, amines, aromatic hydrocarbons (44–46). | Unknown, possibly involving past aqueous environment. | Relevant as a potential source of N-containing organic material on the planetary surface. Organic material can be highly localized and concentrated in the form of "nanoglobules". |
| Comets | High abundance of organics (47). Diverse organic molecules, including N-containing such as glycine, methylamine, ethylamine (48). | Possibly formed from interstellar materials that have been processed in the protosolar nebulae. | Relevant as a potential source of N-containing organic material on the planetary surface. Organic material can be highly concentrated in comet dust particles. |
| Ceres | Large abundance of aliphatic organics (49–52). | Formed endogenously via an unknown process potentially involving past aqueous | Provides a precedent for large abundance of endogenously produced organic molecules (if the |

| | | environment or delivered locally by organic-rich impactors. | hypothesis on endogenous source of organics is confirmed). |
|---|---|---|---|
| Jupiter | Ethane, acetylene, other hydrocarbons (53). | Photochemically generated from methane in the atmosphere. | N/A |
| Ganymede | Various hydrocarbons and possibly aliphatic aldehydes (54). | Endogenous origin, inside Ganymede, possibly hydrothermal. | N/A |
| Callisto | Tholins and other organic molecules containing nitrile groups (55). | UV and particle radiation from the Jovian magnetosphere. | UV- and radiation-driven synthesis of organics could also proceed on the surface of an airless planet. |
| Titan | Large abundance of hydrocarbons and N-rich organic compounds present both as atmospheric hazes and surface deposits (56); Potential presence of large abundance of complex organics in Titan's interior (57, 58). | Photochemistry (solar UV) and atmospheric chemistry converting $CH_4$ and $N_2$ into complex organics; Unknown processes in the interior of Titan, if the interior contains large deposits of complex organics. | If large deposits of diverse organic material in the interior of Titan is confirmed, it suggests unknown endogenous processes that lead to the formation of organics within interiors of planetary bodies. |
| Enceladus | Complex organics (>200 atomic mass units) with diverse functional groups, including aromatics, carbonyls and N-containing molecules (59, 60). | Unknown. Possibly hydrothermal activity in the moon's interior, at the liquid water-rock interface. | N/A |
| Pluto and Charon | Possible presence of hydrocarbons and nitriles that are converted to complex organic molecules on the surface of the dwarf planet (61–64). | Surface ice interactions with solar UV or cosmic rays. | Potential for generation of complex N-containing organics on the surface of the planet via cosmic radiation. |

## S5. List of Abbreviations

### Abbreviations

FTIR spectroscopy: Fourier Transform Infrared Spectroscopy
NMR spectroscopy: Nuclear Magnetic Resonance Spectroscopy
TGA: Thermogravimetric Analysis

## Chemical Nomenclature

The cation names for amino acids (e.g., glycininium, alaninium) reflect their protonated forms, as is standard in amino acid-based ionic liquid naming conventions. The naming of ionic liquids containing nucleic acid bases (thymine, purine, guanine) follows similar convention (e.g., thyminium). For complexes with additional $H_2SO_4$, the neutral molecule is included in the name.

[glycine$^+$][HSO$_4^-$]: Glycininium hydrogen sulfate
[glycine$^+$][HSO$_4^-$][H$_2$SO$_4$]: Glycininium hydrogen sulfate sulfuric acid
[glycine$^+$][HSO$_4^-$][H$_2$SO$_4$]$_2$: Glycininium hydrogen sulfate bis(sulfuric acid)
[histidine$^+$][HSO$_4^-$]: Histidinium hydrogen sulfate
[alanine$^+$][HSO$_4^-$]: Alaninium hydrogen sulfate
[cysteine$^+$][HSO$_4^-$]: Cysteinium hydrogen sulfate
[valine$^+$][HSO$_4^-$]: Valinium hydrogen sulfate
[thymine$^+$][HSO$_4^-$]: Thyminium hydrogen sulfate
[purine$^+$][HSO$_4^-$]: Purinium hydrogen sulfate
[guanine$^+$][HSO$_4^-$]: Guaninium hydrogen sulfate
[1-ethyl-3-methylimidazolium$^+$][HSO$_4^-$]: 1-Ethyl-3-methylimidazolium hydrogen sulfate
[tetrabutylammonium$^+$][HSO$_4^-$]: Tetrabutylammonium hydrogen sulfate
[alanylglycine$^+$][HSO$_4^-$]: Alanylglycinium hydrogen sulfate
[choline$^+$][HSO$_4^-$]: choline hydrogen sulfate

DMSO-$d_6$: dimethyl sulfoxide-$d_6$

## S6. Supplementary Datasets

*Supplementary Dataset S1:* The original $^1$H NMR, $^{31}$P NMR data used in support of the chemical and physical characterization of ionic liquid formation. All data can be downloaded from Zenodo at https://zenodo.org/records/15596466.

*Supplementary Dataset S2:* The original FTIR data used in support of the chemical and physical characterization of ionic liquid formation. All data can be downloaded from Zenodo at https://zenodo.org/records/15596466.

*Supplementary Dataset S3:* The original TGA data used in support of the characterization of ionic liquid formation. All data can be downloaded from Zenodo at https://zenodo.org/records/15596466.

## S7. Supplementary References